\documentclass[english,twocolumn,reprint,aip,jcp]{revtex4-1}
\usepackage[latin9]{inputenc}
\usepackage{amsmath}
\usepackage{amssymb}
\usepackage{graphicx}
\usepackage{babel}
\begin{document}

\title{Inchworm Monte Carlo for exact non-adiabatic dynamics I. Theory and
algorithms}

\author{Hsing-Ta Chen}

\affiliation{Department of Chemistry, Columbia University, New York, New York
10027, U.S.A.}

\affiliation{The Raymond and Beverly Sackler Center for Computational Molecular
and Materials Science, Tel Aviv University, Tel Aviv 69978, Israel}

\author{Guy Cohen}

\affiliation{The Raymond and Beverly Sackler Center for Computational Molecular
and Materials Science, Tel Aviv University, Tel Aviv 69978, Israel}

\affiliation{School of Chemistry, The Sackler Faculty of Exact Sciences, Tel Aviv
University, Tel Aviv 69978, Israel}

\author{David R. Reichman}

\affiliation{Department of Chemistry, Columbia University, New York, New York
10027, U.S.A.}
\begin{abstract}
In this paper we provide a detailed description of the inchworm Monte
Carlo formalism for the exact study of real-time non-adiabatic dynamics.
This method optimally recycles Monte Carlo information from earlier
times to greatly suppress the dynamical sign problem. Using the example
of the spin\textendash boson model, we formulate the inchworm expansion
in two distinct ways: The first with respect to an expansion in the
system\textendash bath coupling and the second as an expansion in
the diabatic coupling. The latter approach motivates the development
of a cumulant version of the inchworm Monte Carlo method, which has
the benefit of improved scaling. This paper deals completely with
methodology, while Paper II provides a comprehensive comparison of
the performance of the inchworm Monte Carlo algorithms to other exact
methodologies as well as a discussion of the relative advantages and
disadvantages of each. 
\end{abstract}
\maketitle

\section{Introduction\label{sec:Introduction}}

The description of real-time dynamics in many-body quantum systems
continues to provide major challenges for current research. An accurate
theoretical understanding of nonequilibrium processes ranging from
charge and energy transport in quantum dots and molecular junctions\cite{Nitzan2003Electrontransportin}
to laser-induced electronic phase transformations\cite{Fausti2011Lightinducedsuperconductivity}
is crucial for the interpretation of experimental results and the
eventual design of new materials and technologies. Quantum Monte Carlo
(QMC) techniques form the basis for the exact description of the \emph{thermodynamics}
of systems dominated by quantum fluctuations.\cite{Negele1988Quantummanyparticle}
In this setting, a variety of QMC methods may be used to \emph{exactly}
calculate the properties of lattice and continuum systems, including
systems where boson particle statistics induce non-trivial collective
phenomena such as the transition to a superfluid state.\cite{Bloch2008Manybodyphysics,Trotzky2010Suppressioncriticaltemperature,Kashurnikov2002Revealingsuperfluidchar21Mottinsulator}
The inclusion of fermionic statistics within QMC is more difficult,
reflecting the NP-hardness of the generic electronic structure problem.\cite{Troyer2005ComputationalComplexityand}
This difficulty reveals itself in the \char`\"{}fermionic sign problem,\char`\"{}
where Monte Carlo summands alternate sign, leading to a poor signal-to-noise
ratio that can inhibit the accurate calculation of the thermodynamic
properties of fermionic assemblies. Despite this difficulty, the umbrella
of QMC techniques has essentially solved the problem of the thermodynamics
of non-fermionic systems,\cite{Shakhnovich1991Proteinfoldingbottlenecks,Saxton1996Anomalousdiffusiondue,Prokofev2001WormAlgorithmsClassical,Newman1999MonteCarloMethods}
while great progress continues to be made towards the development
of accurate QMC approaches for fermions.\cite{Mak1990Solvingsignproblem,Burovski2006CriticalTemperatureand,Hirsch1982MonteCarlosimulations,Hirsch1988Stablemontecarlo,Egger2000PathintegralMonte,Maier2005Quantumclustertheories,Needs2010Continuumvariationaland,LeBlanc2015SolutionsTwoDimensional}

The simulation of real-time quantum dynamics presents another layer
of difficulty that is absent when thermodynamics alone is considered.
In general, when considering the exact simulation of quantum dynamics,
the computational cost scales exponentially with increasing time.
This poor scaling manifests in distinct ways in different methodologies.\cite{Rombouts1998discreteHubbardStratonovich,Muehlbacher2008Realtimepath,Schiro2010Realtimedynamics,Werner2006Continuoustimesolver,Werner2006Hybridizationexpansionimpurity,Gull2008Continuoustimeauxiliary,Gull2010Boldlinediagrammatic,Gull2011ContinuoustimeMonte}
Within attempts to extend QMC to the real-time axis, exponentially
poor scaling arises from the oscillating phase factors generated by
the time evolution operator $e^{-iHt}$. The summation of random phase
information leads to a shrinking signal to noise ratio known as the
\char`\"{}dynamical sign problem\char`\"{}. This afflicts all dynamical
QMC simulations, regardless of the nature of the underlying particle
statistics.\cite{Egger2000PathintegralMonte,Maier2005Quantumclustertheories,Needs2010Continuumvariationaland,LeBlanc2015SolutionsTwoDimensional}

Modern diagrammatic variants of QMC (dQMC) have proven extremely
powerful in the study of thermodynamic properties of impurity models,
which consist of a small interacting subsystem coupled to noninteracting
fermionic or bosonic baths.\cite{Gull2011ContinuoustimeMonte} The
extension of these approaches to real-time dynamics has also met with
some success.\cite{Werner2006Continuoustimesolver,Werner2006Hybridizationexpansionimpurity,Muehlbacher2008Realtimepath,Werner2009DiagrammaticMonteCarlo,Werner2010Dynamicalscreeningin,Schiro2010Realtimedynamics}
In particular, in conjunction with partial resummations of the exact
diagrammatic series\cite{Gull2010Boldlinediagrammatic,Gull2011ContinuoustimeMonte}
and reduced dynamics techniques,\cite{Cohen2011Memoryeffectsin,Cohen2013Generalizedprojecteddynamics}
real-time dQMC has proven capable of the exact simulation of nonequilibrium
properties in the paradigmatic Anderson model for non-trivial time
scales in select parameter regimes.\cite{Cohen2013Numericallyexactlong}
Despite the aforementioned successes, previous real-time dQMC methods
have all been plagued by the dynamical sign problem to differing degrees.\cite{Egger1992QuantumMonteCarlo,Egger1994Lowtemperaturedynamical,Muehlbacher2003Crossoverfromnonadiabatic,Gull2010Boldlinediagrammatic,Cohen2014Greensfunctionsfrom,Cohen2014Greensfunctionsfroma}
Very recently, a new dQMC method, dubbed the \char`\"{}inchworm algorithm,\char`\"{}
has been introduced that largely overcomes the dynamical sign problem.\cite{Cohen2015Tamingdynamicalsign}
The inchworm algorithm optimally recycles diagrammatic information
so that the computational cost scales approximately quadratically,
as opposed to exponentially, with time. For the case of the Anderson
model, the inchworm algorithm has enabled exact real-time simulation
even deep within strongly correlated regions of the parameter space,
such as the Kondo and mixed valence regimes.

While progress for the Anderson model has been impressive, it should
be noted that the number and range of exact benchmarks for this model
are far fewer than those available for a simpler impurity model: the
spin\textendash boson model. The spin\textendash boson model consists
of a two-level system coupled linearly to a bosonic bath, and constitutes
the basic proxy for dissipative condensed phase charge and energy
transfer problems.\cite{Leggett1987Dynamicsdissipativetwo,Weiss1999Quantumdissipativesystems,Nitzan2006ChemicalDynamicsin}
Two decades of numerical effort aimed at the spin\textendash boson
problem have produced a suite of methodologies capable of long-time
simulation of nonequilibrium observables over essentially the entire
parameter space of the model.\cite{Wang2001Systematicconvergencein,Thoss2001Selfconsistenthybrid,Wang2003Multilayerformulationmulticonfiguration,Egger1992QuantumMonteCarlo,Egger1994Lowtemperaturedynamical,Mak2007MonteCarloMethods,Egger2000PathintegralMonte,Makarov1994Pathintegralsdissipative,Makri1995Numericalpathintegral,Tanimura1989TimeEvolutionQuantum,Struempfer2012OpenQuantumDynamics}
In this sense, the spin\textendash boson model embodies a stringent
test which should be passed by any new numerically exact approach
to real-time quantum dynamics.

In the following work, we use the spin\textendash boson model as
a platform to provide the essential details of the inchworm approach
and to improve and expand upon the methodology. In particular we describe
two diagrammatic expansions (and their resummations within the inchworm
framework) rooted in distinct exactly solvable reference systems.
We further introduce a new cumulant-based approach\cite{VanKampen1974cumulantexpansionstochastic,Yoon1975comparisongeneralizedcumulant,Mukamel1979NonMarkoviantheory,Reichman1997Cumulantexpansionsand}
that reduces the computational cost from quadratic to linear in time.
In essence, the use of cumulants allows for the construction of an
inchworm expansion for the memory function directly from the moment
expansion and without the need for any \emph{a priori} information
about the memory kernel itself. We argue that taken together, the
distinct inchworm algorithms presented here should essentially cover
the relevant parameter space of the spin\textendash boson model. We
defer the detailed comparison of our new approach to established benchmarks,
as well as a discussion of the relative benefits and drawbacks of
our approach, to a companion paper.

We concentrate on a minimal model in equilibrium here, but expect
most of the impact of this methodology to come from generalizations.
The generalization to nonequilibrium scenarios, such as multiple baths
and time-dependent Hamiltonians, is trivial and will incur no meaningful
computational cost. The generalization to multilevel systems is also
straightforward: for the system\textendash bath coupling expansion,
the rank of the propagator matrix is the number of subsystem states
$N$, and the algorithm should scale as the cost of matrix multiplications,
at $O\left(N^{3}\right)$. In practice, the propagator matrix may
be sparse under certain conditions, such that better scaling may be
achievable. On the other hand, the diabatic coupling expansion is
very promising in that it can be formulated entirely in the language
of single-particle Green's functions rather than many-body states.
In this form, it should scale as $O\left(\left(\log N\right)^{3}\right)$,
similarly to certain equilibrium methods.\cite{Gull2008Continuoustimeauxiliary,Gull2011ContinuoustimeMonte}

The organization of the paper is as follows. In Sec.~\ref{sec:dQMC-scheme-and},
we review the real-time dQMC scheme and the inchworm algorithm in
a general formalism. In Sec.~\ref{sec:System=002013Bath-Coupling-Inchworm},
we formulate the system\textendash bath coupling expansion and its
corresponding inchworm expansion. In Sec.~\ref{sec:Diabatic-Coupling-Expansion},
the diabatic coupling expansion is described. In Sec.~\ref{sec:Diabatic-Coupling-Cumulant},
we introduce cumulant inchworm expansions based on the diabatic coupling
expansion. Our conclusions are presented in Sec.~\ref{sec:Conclusions}.

\section{dQMC scheme and the inchworm algorithm\label{sec:dQMC-scheme-and}}

In this section, we briefly review the real-time dQMC approach,\cite{Gull2011ContinuoustimeMonte}
the emergence of the dynamical sign problem and the inchworm algorithm\cite{Cohen2015Tamingdynamicalsign}
in a general framework.

We consider a generic Hamiltonian of an open quantum system in the
form
\begin{equation}
H=H_{s}+H_{b}+H_{sb},\label{eq:system-bath_Hamiltonian}
\end{equation}
where $H_{s}$ and $H_{b}$ are the Hamiltonian of the system and
the bath, respectively, and $H_{sb}$ describes the system\textendash bath
coupling. For a given observable $O$, we are interested in its time-dependent
expectation value
\begin{equation}
\left\langle O\left(t\right)\right\rangle =\mathrm{Tr}\left\{ \rho_{0}e^{iHt}Oe^{-iHt}\right\} .\label{eq:observable}
\end{equation}
Here, $\left\langle \cdot\right\rangle =\mathrm{Tr}\left\{ \rho_{0}\cdot\right\} $
is the trace performed over all degrees of freedom and $\rho_{0}$
is the initial density matrix of the full system. It should be noted
that equilibrium time correlation functions may also be calculated
within the framework outlined below,\cite{Cohen2014Greensfunctionsfrom,Cohen2013Numericallyexactlong}
however for simplicity we focus on one-time non-equilibrium quantities
of the form (\ref{eq:observable}).

\subsection{Dyson series}

To evaluate the dynamics of the observable $\left\langle O\left(t\right)\right\rangle $,
a key needed element is the propagator $e^{-iHt}$, which is difficult
to calculate in a computationally useful form. In general, we can
expand the propagator in a perturbative fashion by writing the Hamiltonian
as 
\begin{eqnarray}
H & = & H_{0}+H^{\prime},\label{eq:perturbative_hamiltonian}
\end{eqnarray}
thus partitioning $H$ into a (solvable) $H_{0}$ and an interaction
Hamiltonian $H^{\prime}$. In this interaction picture, the dynamics
of an operator $O$ is given by 
\begin{equation}
e^{iHt}Oe^{-iHt}=U^{\dagger}\left(t\right)\widetilde{O}\left(t\right)U\left(t\right),\label{eq:dynamics}
\end{equation}
where the propagator is $U\left(t\right)$ given by $U\left(t\right)=e^{iH_{0}t}e^{-iHt}$.
We denote the time-dependent operator in the interaction picture by
$\widetilde{O}\left(t\right)=e^{iH_{0}t}Oe^{-iH_{0}t}$. One can expand
the propagator using the time-ordered Dyson series ($\hbar=1$) 
\begin{equation}
\begin{split}U\left(t\right)= & \sum_{n=0}^{\infty}\left(-i\right)^{n}\int_{0}^{t}\mathrm{d}t_{1}\int_{0}^{t_{1}}\mathrm{d}t_{2}\cdots\int_{0}^{t_{n-1}}\mathrm{d}t_{n}\\
 & ~\times\widetilde{H}^{\prime}\left(t_{1}\right)\widetilde{H}^{\prime}\left(t_{2}\right)\cdots\widetilde{H}^{\prime}\left(t_{n}\right)
\end{split}
\end{equation}
which contains a series of interaction operators $\widetilde{H}^{\prime}\left(t_{i}\right)=e^{iH_{0}t}H^{\prime}e^{-iH_{0}t}$
with the chronological time ordering $t>t_{1}>t_{2}>\cdots>t_{n}>0$.
If this expansion is applied to the two interaction picture propagators
in Eq.~\ref{eq:dynamics}, the folded Keldysh contour naturally emerges
from the sequence of interaction operators generated by the product.
The interaction operators arising from $U\left(t\right)$ have time
arguments denoted as $\left\{ t_{i}^{+}\right\} $, and are thought
of as existing on the forward or $+$ branch of the contour, while
those emanating from $U^{\dagger}\left(t\right)$ have time arguments
denoted as $\left\{ t_{i}^{-}\right\} $, and exist on the backward
or $-$ branch. This is illustrated in Fig.~\ref{fig:Keldysh-times}a.
The contour is folded at $t=t_{\mathrm{max}}$, where the observable
operator is applied. Each set of time arguments, $\left\{ t_{i}^{+}\right\} $
and $\left\{ t_{i}^{-}\right\} $, is time ordered: \textbf{$t_{\mathrm{max}}>t_{1}^{\pm}>t_{2}^{\pm}>\cdots>0^{\pm}$},
where $0^{\pm}$ denote the initial time on the $\pm$ branch, respectively.
Therefore, we can write Eq.~\ref{eq:dynamics} as
\begin{equation}
\begin{split}O\left(t\right)= & \sum_{n=0}^{\infty}\int_{0}^{t_{\mathrm{max}}}\mathrm{d}t_{1}^{+}\int_{0}^{t_{1}^{+}}\mathrm{d}t_{2}^{+}...\int_{0}^{t_{n-1}^{+}}\mathrm{d}t_{n}^{+}\times\\
 & \sum_{n'=0}^{\infty}\int_{0}^{t_{\mathrm{max}}}\mathrm{d}t_{1}^{-}\int_{0}^{t_{1}^{-}}\mathrm{d}t_{2}^{-}...\int_{0}^{t_{n^{\prime}-1}^{-}}\mathrm{d}t_{n^{\prime}}^{-}\times\\
 & \ \left(-i\right)^{n}i^{n^{\prime}}\widetilde{H}^{\prime}\left(t_{n'}^{-}\right)\cdots\widetilde{H}^{\prime}\left(t_{1}^{-}\right)\times\\
 & \ \widetilde{O}\left(t_{\mathrm{max}}\right)\widetilde{H}^{\prime}\left(t_{1}^{+}\right)\cdots\widetilde{H}^{\prime}\left(t_{n}^{+}\right).
\end{split}
\end{equation}

For brevity, it will be convenient to write the two types of time
arguments on the two branches of the contour in terms of a single
time argument label $s_{i}$:
\begin{equation}
s_{i}=\begin{cases}
s_{i}^{+}=t_{n-i+1}^{+} & i\le n,\\
s_{i}^{-}=t_{i-n}^{-} & n<i\le m.
\end{cases}\label{eq:contour_time}
\end{equation}
Here, $m=n+n^{\prime}$ and $\left\{ s_{i}\right\} $ is ordered according
to the Keldysh contour causality, $s_{1}<\cdots<s_{m}$ as shown in
Fig.~\ref{fig:Keldysh-times}. We define $s_{i}<s_{j}$ if $s_{i}$
occurs before $s_{j}$ on the Keldysh contour. Therefore, we can write
Eq.~\ref{eq:dynamics} as an expansion in terms of $s_{i}$,
\begin{equation}
\begin{split}O\left(t\right)= & \sum_{m=0}^{\infty}\sum_{n=0}^{m}\int\mathrm{d}s_{m}\cdots\int\mathrm{d}s_{1}\left(-1\right)^{n}i^{m}\times\\
 & \widetilde{H}^{\prime}\left(s_{m}\right)\cdots\widetilde{H}^{\prime}\left(s_{n+1}\right)\widetilde{O}\left(t_{\mathrm{max}}\right)\widetilde{H}^{\prime}\left(s_{n}\right)\cdots\widetilde{H}^{\prime}\left(s_{1}\right),
\end{split}
\end{equation}
where the integration $\int\mathrm{d}s_{m}\cdots\int\mathrm{d}s_{1}$
is taken to represent
\begin{equation}
\begin{split}\int\mathrm{d}s_{m}\cdots\int\mathrm{d}s_{1}= & \int_{0}^{t_{\mathrm{max}}}\mathrm{d}t_{1}^{+}\int_{0}^{t_{1}^{+}}\mathrm{d}t_{2}^{+}...\int_{0}^{t_{n-1}^{+}}\mathrm{d}t_{n}^{+}\\
 & ~\int_{0}^{t_{\mathrm{max}}}\mathrm{d}t_{1}^{-}\int_{0}^{t_{1}^{-}}\mathrm{d}t_{2}^{-}...\int_{0}^{t_{n^{\prime}-1}^{-}}\mathrm{d}t_{n^{\prime}}^{-}.
\end{split}
\end{equation}
Each term in the expansion can be represented by diagrams, in which
a vertex or open circle in Fig.~1(a) represents the interactions
occurring at the times $\left\{ s_{i}\right\} $ and a cross symbol
indicates the tip or the folding time $t_{\mathrm{max}}$ of the Keldysh
contour where the observable operator acts. For instance, Fig.~\ref{fig:Keldysh-times}b
shows the diagrams of the unperturbed term ($m=0$) and some example
diagrams of second order ($m=2$, two vertices) and of fourth order
($m=4$, four vertices).
\begin{center}
\begin{figure}
\begin{raggedright}
(a)
\par\end{raggedright}
\begin{centering}
\includegraphics{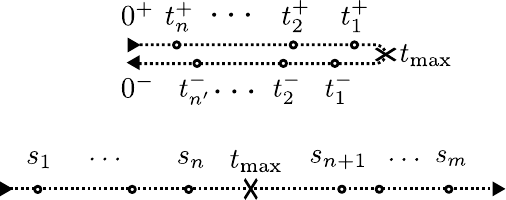}
\par\end{centering}
\begin{raggedright}
(b)
\par\end{raggedright}
\begin{centering}
\includegraphics{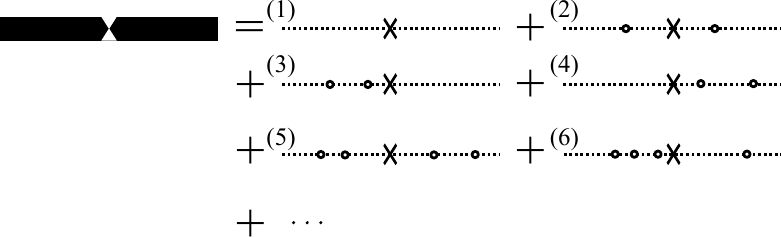}
\par\end{centering}
\caption{(a) A configuration $\boldsymbol{s}$ drawn on the Keldysh contour,
with physical times $t_{i}$ on the forward or $+$ branch and $t_{i}^{\prime}$
on the backward or $-$ branch. Below, the configuration is shown
on the unfolded contour with contour times $s_{i}$. The $\boldsymbol{\times}$
indicates the tip or fold of the contour and the ticks indicate interaction
operators $H^{\prime}$. (b) General framework of \emph{bare} dQMC.
The thin line represents an unperturbed propagator $e^{-iH_{0}s}$,
while the thick line represents the exact sum over all possible configurations
contributing to some observable $\left\langle O\left(t\right)\right\rangle $.
(1) is the zeroth ($m=0$) order contribution, $\Bigl\langle e^{iH_{0}t}Oe^{-iH_{0}t}\Bigr\rangle$.
(2)\textendash (4) are examples of second ($m=2$) order contributions
with (2)$n=1$, (3)$n=2$, and (4)$n=0$. (5) and (6) are examples
of fourth ($m=4$) order configurations.\label{fig:Keldysh-times}}
\end{figure}
\par\end{center}

\subsection{Real-time path integral formulation}

The dynamical quantities of interest can be expressed in the form
of a path integral, or more generally the integral over the contour
configuration space

\begin{equation}
\left\langle O\left(t\right)\right\rangle =\int\mathrm{d}\boldsymbol{s}{\cal O}\left(\boldsymbol{s}\right),\label{eq:path_integral_1}
\end{equation}
where we denote $\boldsymbol{s}=\left\{ s_{i}\right\} $ as the contour
configuration. Note that this expression is implicitly time-ordered
and the integration $\int\mathrm{d}\mathbf{s}$ is taken to mean
\begin{equation}
\int\mathrm{d}\boldsymbol{s}=\sum_{m=0}^{\infty}\sum_{n=0}^{m}\int\mathrm{d}s_{m}\cdots\int\mathrm{d}s_{1}.
\end{equation}
The contribution of a given configuration $\boldsymbol{s}$ is given
by 
\begin{equation}
\begin{split}{\cal O}\left(\boldsymbol{s}\right)=\left(-1\right)^{n}i^{m} & \Bigl\langle\widetilde{H}^{\prime}\left(s_{m}\right)\cdots\widetilde{H}^{\prime}\left(s_{n+1}\right)\times\\
 & \widetilde{O}\left(t_{\mathrm{max}}\right)\widetilde{H}^{\prime}\left(s_{n}\right)\cdots\widetilde{H}^{\prime}\left(s_{1}\right)\Bigr\rangle.
\end{split}
\label{eq:functional_general}
\end{equation}
This object is a seemingly complicated multi-time quantity, but in
many cases is that it can be efficiently evaluated since it is defined
by an interaction picture under the propagation associated with a
solvable $H_{0}$. 

The dQMC method provides an unbiased estimator for the infinite-dimensional
integral over all configuration parameters, $\int\mathrm{d}\boldsymbol{s}{\cal O}\left(\boldsymbol{s}\right)$,
by summing over a set of sample configurations $\boldsymbol{s}_{i}$
drawn from some normalized probability distribution defined by $\mathrm{Prob}\left(\boldsymbol{s}\right)=\frac{\mathrm{w}\left(\boldsymbol{s}\right)}{\int\mathrm{d}\boldsymbol{s}\mathrm{w}\left(\boldsymbol{s}\right)}\equiv\frac{\mathrm{w}\left(\boldsymbol{s}\right)}{Z_{\mathrm{w}}}$.
The Metropolis\textendash Hastings algorithm\cite{Metropolis1953EquationStateCalculations,Hastings1970MonteCarlosampling}
is method for generating a sample set of this type when only $\mathrm{w}\left(\boldsymbol{s}\right)$
is known. To see how this is used, consider that for any prescribed
weight function $\mathrm{w}\left(\boldsymbol{s}\right)$, we have
\begin{equation}
\int\mathrm{d}\boldsymbol{s}{\cal O}\left(\boldsymbol{s}\right)=Z_{\mathrm{w}}\int\mathrm{d}\boldsymbol{s}\frac{{\cal O}\left(\boldsymbol{s}\right)}{\mathrm{w}\left(\boldsymbol{s}\right)}\mathrm{Prob}\left(\boldsymbol{s}\right).
\end{equation}
Given that the $\left\{ \boldsymbol{s}_{i}\right\} $ for $i\in\left\{ 1,...,M\right\} $
are drawn from $\mathrm{Prob}\left(\boldsymbol{s}\right)$, in the
limit of large $M$ one obtains
\begin{equation}
\int\mathrm{d}\boldsymbol{s}{\cal O}\left(\boldsymbol{s}\right)\simeq\frac{Z_{\mathrm{w}}}{M}\sum_{i=1}^{M}\frac{{\cal O}\left(\boldsymbol{s}_{i}\right)}{\mathrm{w}\left(\boldsymbol{s}_{i}\right)}\equiv Z_{\mathrm{w}}\left\langle \frac{{\cal O}}{\mathrm{w}}\right\rangle _{\mathrm{w}}.
\end{equation}
Importantly, we note that $Z_{\mathrm{w}}$ is completely independent
of the observable calculated. Therefore, to remove the dependence
on $Z_{\mathrm{w}}$, we introduce a ``normalizing'' observable
$N=\int\mathrm{d}\boldsymbol{s}\mathcal{N}\left(\boldsymbol{s}\right)$
which can be evaluated analytically. Evaluating $N$ via the same
Monte Carlo procedure, one obtains
\begin{equation}
N=\int\mathrm{d}\boldsymbol{s}\mathcal{N}\left(\boldsymbol{s}\right)\simeq Z_{\mathrm{w}}\left\langle \frac{{\cal N}}{\mathrm{w}}\right\rangle _{\mathrm{w}}.
\end{equation}
With this normalization, $Z_{\mathrm{w}}$ cancels out of all final
expressions:
\begin{equation}
\int\mathrm{d}\boldsymbol{s}{\cal O}\left(\boldsymbol{s}\right)=N\frac{\int\mathrm{d}\boldsymbol{s}{\cal O}\left(\boldsymbol{s}\right)}{\int\mathrm{d}\boldsymbol{s}{\cal N}\left(\boldsymbol{s}\right)}\simeq N\frac{Z_{\mathrm{w}}\left\langle \frac{{\cal O}}{\mathrm{w}}\right\rangle _{\mathrm{w}}}{Z_{\mathrm{w}}\left\langle \frac{{\cal N}}{\mathrm{w}}\right\rangle _{\mathrm{w}}}=N\frac{\left\langle \frac{{\cal O}}{\mathrm{w}}\right\rangle _{\mathrm{w}}}{\left\langle \frac{{\cal N}}{\mathrm{w}}\right\rangle _{\mathrm{w}}}.
\end{equation}
Since we have complete freedom in the choice of $N$, one is free
to choose a quantity which is easy to evaluate in both the Monte Carlo
and the analytical calculation. The choice used here is $\mathcal{N}\left(\boldsymbol{s}\right)=1$,
such that $N$ is simply the hypervolume of the multidimensional space
of interaction times. Since this hypervolume normalization is positive
definite, it cannot have a sign problem, and all potential sign problems
must appear in the nominator. For $\mathrm{w}\left(\boldsymbol{s}\right)$,
we typically choose the absolute value $\left|\mathcal{O}\left(\boldsymbol{s}\right)\right|$
of the contribution to the observable itself or a closely related
property, such that the summation is optimized for summing large contributions
to a particular observable. It is currently unknown whether this choice
is optimal. One may choose $\mathrm{w}\left(\boldsymbol{s}\right)$
as any positive definite functional of the configuration $\boldsymbol{s}$
does not affect the correctness and accuracy of the algorithm. Yet,
different choices of $\mathrm{w}\left(\boldsymbol{s}\right)$ can
strongly affect the precision and efficiency. For a given amount of
computer time, one will be left with different error estimates. 

\subsection{Dynamical sign problem and inchworm algorithms}

Unfortunately, summing individual contributions to an observable
in this manner, the so-called \emph{bare} dQMC algorithm, generally
involves a dynamical sign problem. In real-time dQMC, the dynamical
sign problem is caused by the oscillatory nature of real-time propagators
which results in exponentially growing computational cost as time
increases.\cite{Muehlbacher2008Realtimepath,Werner2009DiagrammaticMonteCarlo,Schiro2010Realtimedynamics,Antipov2016VoltageQuenchDynamics}
To circumvent the dynamical sign problem, we employ inchworm expansions.\cite{Cohen2015Tamingdynamicalsign}
This allows us to efficiently reuse quantities propagated within short
time intervals in the calculation of quantities propagated between
longer times. Two concrete examples of practical inchworm algorithms
for the spin\textendash boson model will be developed below.

We briefly introduce the general concept behind inchworm expansions.
Let $s_{i}<s_{\uparrow}<s_{f}$ be three times: an ``initial,''
``inchworm'' and ``final'' time, respectively. Assume some set
of properties have been exactly evaluated for all cases where all
interaction vertices are restricted to the time interval $\left[s_{i},s_{\uparrow}\right]$.
Given knowledge of these auxiliary restricted quantities, it is often
possible to construct an efficient expansion for the same set of quantities
with the vertices restricted to the \emph{longer} interval $\left[s_{i},s_{f}\right]$.
This describes an \emph{inchworm step}, or the process of \emph{inching}.
A series of inchworm steps allows one to start with a set of easily
evaluated restricted quantities defined over very short intervals,
eventually obtaining the set of unrestricted physical quantities for
which interaction vertices span the full length of the Keldysh contour.

The inchworm algorithm has the distinct advantage (when compared
to its bare equivalent) that much fewer diagrams must be sampled to
obtain a converged answer, since each inchworm diagram contains an
infinite number of bare diagrams. Often, relatively few low-order
inchworm diagrams contain all important contributions from the relevant
bare diagrams at all orders. This advantage comes at two important
costs. First, as when working with nonequilibrium Green's functions,
one is forced to calculate a complete set of two-time properties even
if only single-time properties are of interest. Specifically, all
propagators between any two points in $\left[s_{i},s_{\uparrow}\right]$
are required to obtain a propagator between $s_{i}$ and $s_{f}$.
The memory cost of the algorithm is proportional to the square of
the number of points on the time contour. For a naive implementation
(where this information is neither distributed nor shared) that data
is duplicated to all processes in the final stages of the calculation,
rapidly becoming a limiting factor. Second, Monte Carlo evaluations
at long times are no longer independent of short-time evaluations,
and errors are carried forward in time during the stepping procedure.
This has profound computational implications in that the algorithm
is not ``embarrassingly parallel'' like standard Monte Carlo techniques,
since information concerning short-time propagators must be distributed
between the various computer nodes performing the calculation. Furthermore,
careful error analysis is required in order to take error propagation
into account. Essentially, a series of completely independent calculations
must be carried out to evaluate the statistical errors, and one must
then verify that systematic errors due to the error propagation (in
addition to the statistical ones common to all Monte Carlo techniques)
are assessed and converged to within the desired accuracy.\footnote{In our calculations in the companion paper, we compare the full error
estimation and the statistical errors from individual runs, showing
that they follow a nontrivial relationship. It is possible to perform
a full error analysis within a single Monte Carlo run by way of a
recursive bootstrapping procedure; however, this has not been implemented
so far, and it is not clear that it offers an advantage.}

Within the formulation of the inchworm algorithm, each single inchworm
step is numerically exact, in the sense that unbiased results are
obtained with only statistical errors that can be converged to any
desired accuracy by increasing the number of Monte Carlo samples.
The sequence/grid of inchworm steps with finite size becomes exact
at the limit of small time discretization, where all propagators between
shorter time intervals can be interpolated into a smooth functional
form which accurately represents the exact continuous propagator.
In practice, it is usually also necessary to truncate (and converge
in) the maximum order of sampled configuration for each inchworm step.
A more detailed discussion and a sequence of tests will be presented
in the companion paper.

\subsection{Spin\textendash boson model }

We now specialize the discussion to the case of the spin\textendash boson
model. This allows us to give explicit expressions for the terms that
emerge in expansions that employ different choices of $H_{0}$. The
form of the Hamiltonian is given by Eq.~(\ref{eq:system-bath_Hamiltonian}).
The system Hamiltonian $H_{s}$ is taken to be a two-level system
in the diabatic basis $\left|\text{\ensuremath{\alpha}}\right\rangle \in\left\{ \left|1\right\rangle ,\left|2\right\rangle \right\} $,
\begin{equation}
H_{s}=\epsilon\hat{\sigma}_{z}+\Delta\hat{\sigma}_{x}.\label{eq:system_Hamiltonian}
\end{equation}
In this notation, $\hat{\sigma}_{z}=\left|1\right\rangle \left\langle 1\right|-\left|2\right\rangle \left\langle 2\right|$
and $\hat{\sigma}_{x}=\left|1\right\rangle \left\langle 2\right|+\left|2\right\rangle \left\langle 1\right|$.
The energetic bias $\epsilon$ is the energy difference between the
two diabatic states, and the diabatic coupling $\Delta$ characterizes
spin flip processes within the electronic system. The boson bath consists
of a set of harmonic oscillators with frequencies $\omega_{\ell}$
described by the bath Hamiltonian
\begin{equation}
H_{b}=\sum_{\ell}\frac{1}{2}\left(p_{\ell}^{2}+\omega_{\ell}^{2}x_{\ell}^{2}\right)=\sum_{\ell}\omega_{\ell}\left(b_{\ell}^{\dagger}b_{\ell}+\frac{1}{2}\right).\label{eq:bath_Hamiltonian}
\end{equation}
The system\textendash bath coupling $H_{sb}$ is assumed to be linear
in the bath coordinates
\begin{equation}
H_{sb}=\hat{\sigma}_{z}\sum_{\ell}c_{\ell}x_{\ell}.\label{eq:system_bath_coupling}
\end{equation}
The coupling constants $c_{\ell}$ describe the strength of the interaction
between the harmonic modes and the spin. The system\textendash bath
coupling is typically parametrized in compact form by the spectral
density 
\begin{equation}
J\left(\omega\right)=\frac{\pi}{2}\sum_{\ell}\frac{c_{\ell}^{2}}{\omega_{\ell}}\delta\left(\omega-\omega_{\ell}\right).
\end{equation}
Throughout this work, we will concentrate on the local dynamics of
the spin $\hat{\sigma}_{z}$ in the diabatic basis
\begin{equation}
\left\langle \sigma_{z}\left(t\right)\right\rangle =\mathrm{Tr}\left\{ \rho_{0}e^{iHt}\hat{\sigma}_{z}e^{-iHt}\right\} .
\end{equation}
Here we only address factorized initial conditions corresponding
to thermal equilibrium of the bath in the absence of the system\textendash bath
coupling, such that the initial density matrix is given by the factorized
form $\rho_{0}=\rho_{s}\otimes\rho_{b}$, with the bath initially
in equilibrium $\rho_{b}=\frac{e^{-\beta H_{b}}}{\mathrm{Tr}_{b}\left\{ e^{-\beta H_{b}}\right\} }$.
We specify the initial condition of the spin as $\rho_{s}=\left|1\right\rangle \left\langle 1\right|$.
Treatment of more general initial conditions is simple but will not
be discussed further here.

There are several useful ways of partitioning $H$ into $H_{0}$
and $H^{\prime}$ such that the perturbation series of Eq.~\ref{eq:perturbative_hamiltonian}
can be carried out, each yielding a different type of expansion. We
will discuss two such choices. One treatment takes $H^{\prime}=H_{sb}$,
expanding with respect to the system\textendash bath coupling. Another
takes $H^{\prime}=\Delta\hat{\sigma}_{x}$, expanding in the diabatic
coupling $\Delta$. In Secs.~\ref{sec:System=002013Bath-Coupling-Inchworm}\textendash \ref{sec:Diabatic-Coupling-Cumulant},
we discuss these expansions and their inchworm Monte Carlo implementations.

\section{System\textendash Bath Coupling Inchworm (SBCI) Expansion\label{sec:System=002013Bath-Coupling-Inchworm}}

\subsection{Bare dQMC}

We start with the example of the bare dQMC expansion in terms of
the system\textendash bath coupling $H^{\prime}=H_{sb}$. This expansion
is analogous to the hybridization expansion in the Anderson model,
for which the first inchworm expansion was formulated. The unperturbed
Hamiltonian is taken to be $H_{0}=H_{s}+H_{b}$ and the initial density
matrix is $\rho_{0}=\left|1\right\rangle \left\langle 1\right|\otimes\rho_{b}$.
To write a dQMC expression for the expectation value of the observable
$O=\hat{\sigma}_{z}$, we must determine the contribution ${\cal O}\left(\boldsymbol{s}\right)$
of any given configuration $\boldsymbol{s}$ to this expectation value
in the form of Eq.~\ref{eq:functional_general}:
\begin{equation}
\begin{split}{\cal O}\left(\boldsymbol{s}\right)=\left(-1\right)^{n}i^{m} & \Bigl\langle\widetilde{H}_{sb}\left(s_{m}\right)\cdots\widetilde{H}_{sb}\left(s_{n+1}\right)\times\\
 & \widetilde{\sigma}_{z}\left(t\right)\widetilde{H}_{sb}\left(s_{n}\right)\cdots\widetilde{H}_{sb}\left(s_{1}\right)\Bigr\rangle.
\end{split}
\label{eq:hyb_influence_functional_1}
\end{equation}
In the interaction picture $\widetilde{H}_{sb}\left(s\right)=e^{iH_{0}s}H_{sb}e^{-iH_{0}s}$
can be factorized as
\begin{equation}
\widetilde{H}_{sb}\left(s\right)=\widetilde{\sigma}_{z}\left(s\right)\times\sum_{\ell}c_{\ell}\widetilde{x}_{\ell}\left(s\right),
\end{equation}
and we define the operator of the bath part as
\begin{equation}
\widetilde{B}\left(s\right)=\sum_{k}c_{\ell}\widetilde{x}_{\ell}\left(s\right).
\end{equation}
It turns out that for a linear coupling of the form of Eq.~(\ref{eq:system_bath_coupling}),
one can write Eq.~(\ref{eq:hyb_influence_functional_1}) as a product
of a system influence functional ${\cal U}\left(\boldsymbol{s}\right)$
and a bath influence functional ${\cal L}\left(\boldsymbol{s}\right)$:
\begin{equation}
{\cal O}\left(\boldsymbol{s}\right)=\left(-1\right)^{n}i^{m}{\cal U}\left(\boldsymbol{s}\right){\cal L}\left(\boldsymbol{s}\right).\label{eq:hyb_influence_functional_2}
\end{equation}

The system influence functional ${\cal U}\left(\boldsymbol{s}\right)$
for the given initial condition $\left|1\right\rangle \left\langle 1\right|$
is defined as
\begin{equation}
\begin{split}{\cal U}\left(\boldsymbol{s}\right)= & \left\langle 1\right|\widetilde{\sigma}_{z}\left(s_{m}\right)\cdots\widetilde{\sigma}_{z}\left(s_{n+1}\right)\times\\
 & ~\widetilde{\sigma}_{z}\left(t_{\mathrm{max}}\right)\widetilde{\sigma}_{z}\left(s_{n}\right)\cdots\widetilde{\sigma}_{z}\left(s_{1}\right)\left|1\right\rangle .
\end{split}
\label{eq:matrix_product}
\end{equation}
 For the spin-$\frac{1}{2}$ case, all operators can be written in
the form of matrices of rank 2 in the basis of the Hilbert space of
the isolated spin. Eq.~(\ref{eq:matrix_product}) can then be efficiently
evaluated as a matrix product of unperturbed system propagators of
the form $e^{-iH_{s}\left(s_{i}-s_{i-1}\right)}$, sandwiched between
$\hat{\sigma}_{z}$ operators with $s_{i}-s_{j}$ denoting the difference
of physical times given by Eq.~(\ref{eq:contour_time}).

The bath influence functional is given by an $m$-time interaction
picture correlation function of the bath operator $\widetilde{B}\left(s\right)$
in the form of
\begin{equation}
{\cal L}\left(\boldsymbol{s}\right)=\Bigl\langle\widetilde{B}\left(s_{m}\right)\cdots\widetilde{B}\left(s_{1}\right)\Bigr\rangle_{b},
\end{equation}
where we denote $\left\langle \cdot\right\rangle _{b}=\mathrm{Tr}_{b}\left\{ \rho_{b}\cdot\right\} $
and $\rho_{b}$ is the initial bath density matrix. Using Wick's theorem,\cite{Fetter2003QuantumTheoryMany,Negele1988Quantummanyparticle}
which is valid for the bath operators within the interaction picture,
one can express $\mathcal{L}\left(\mathbf{s}\right)$ as a sum of
products of two-time correlation functions by use of the identity
\begin{equation}
\Bigl\langle\widetilde{B}\left(s_{m}\right)\cdots\widetilde{B}\left(s_{1}\right)\Bigr\rangle_{b}=\sum_{q\in{\cal Q}_{m}}\prod_{(j,k)\in q}\left\langle \widetilde{B}\left(s_{k}\right)\widetilde{B}\left(s_{j}\right)\right\rangle _{b}.\label{eq:hybridization_functional}
\end{equation}
The bath influence functional is zero for odd $m$. ${\cal Q}_{m}$
denotes the set of possible distinct pairings of the integers 1, 2,
..., $m$: each element $q\in\mathcal{Q}_{m}$ is a set of ordered
tuples corresponding to a single pairing. For example, for $m=2$
there is only one pairing, $q=\left\{ \left(2,1\right)\right\} $,
and
\[
{\cal Q}_{2}=\left\{ \left\{ \left(1,2\right)\right\} \right\} .
\]
For $m=4$ there are three possible pairings:
\[
{\cal Q}_{4}=\left\{ \left\{ \left(1,2\right),\left(3,4\right)\right\} ,\left\{ \left(1,3\right),\left(2,4\right)\right\} ,\left\{ \left(1,4\right),\left(2,3\right)\right\} \right\} ,
\]
and so on. With these definitions, the bath influence functional takes
the form
\begin{equation}
{\cal L}\left(\boldsymbol{s}\right)=\sum_{q\in{\cal Q}_{m}}{\cal L}_{q}\left(\boldsymbol{s}\right),\label{eq:bath_functional_sum}
\end{equation}
where the functional of a given pairing $q$;
\begin{equation}
{\cal L}_{q}\left(\boldsymbol{s}\right)=\prod_{(j,k)\in q}\left\langle \widetilde{B}\left(s_{k}\right)\widetilde{B}\left(s_{j}\right)\right\rangle _{b},\label{eq:bath_functional_per_term}
\end{equation}
corresponding to a particular \emph{diagram} with the coupling lines
connecting $s_{j}$ and $s_{k}$ on the Keldysh contour (see Fig.~\ref{fig:hybridization_diagram_4th}b).
The two-time correlation function of the harmonic bath in the interaction
picture can be evaluated semi-analytically prior to the start of the
dQMC calculation as
\begin{equation}
\begin{split} & \left\langle \widetilde{B}\left(s_{k}\right)\widetilde{B}\left(s_{j}\right)\right\rangle _{b}=\frac{2}{\pi}\int d\omega J\left(\omega\right)\times\\
 & \ \left[\coth\left(\frac{\beta\omega}{2}\right)\cos\omega\left(s_{k}-s_{j}\right)-i\sin\omega\left(s_{k}-s_{j}\right)\right].
\end{split}
\end{equation}
In practice, an $m$-time path configuration includes $(m-1)!!$ diagrams,
and computing each diagram requires $m/2$ evaluations of the bath
correlation function. Thus, calculating an $m$\textendash time correlation
function requires a total of $\left(m-1\right)!!\left(m/2\right)$
function evaluation, which approaches $\frac{m}{\sqrt{2}}\left(m/e\right)^{m/2}$
in the large $m$ limit. This rapidly becomes a bottleneck for high
perturbation order. However, rather than explicitly summing over all
diagrams in a configuration, it is possible to sum over the pairings
as defined in Eq.~(\ref{eq:bath_functional_sum}) within the Monte
Carlo procedure, thus effectively removing this scaling issue at the
cost of an overall increase in the sign problem (since the configuration
space is expanded). In practice, this should only be done if extremely
high orders are needed, and it was not necessary anywhere in the companion
paper.

\subsection{Restricted propagators and observables}

To facilitate our discussion of the inchworm algorithm, we now define
\emph{restricted} \emph{propagators} on contour subintervals. Propagators
are defined with respect to particular physical observables. The \emph{bare}
restricted propagator $G_{\alpha\beta}^{\left(0\right)}\left(s_{f},s_{i}\right)$
is defined as follows. When the subinterval $\left[s_{i},s_{f}\right]$
is on a single branch of the contour, such that $s_{i}^{+},s_{f}^{+}<t_{\mathrm{max}}$
or $s_{i}^{-},s_{f}^{-}>t_{\mathrm{max}}$ , then
\begin{equation}
G_{\alpha\beta}^{\left(0\right)}\left(s_{f}^{\pm},s_{i}^{\pm}\right)=\left\langle \alpha\right|e^{-iH_{s}\left(s_{f}^{\pm}-s_{i}^{\pm}\right)}\left|\beta\right\rangle .\label{eq:bare_restricted_prop_def}
\end{equation}
When the endpoints of the interval are on two different branches,
it is defined differently in order to account for the observable at
the contour's folding point. In this case the operator associated
with the observable is $\sigma_{z}\left(t_{\mathrm{max}}\right)$,
such that:
\begin{equation}
G_{\alpha\beta}^{\left(0\right)}\left(s_{f}^{-},s_{i}^{+}\right)=\left\langle \alpha\right|e^{-iH_{s}\left(s_{f}^{-}-t_{\mathrm{max}}\right)}\sigma_{z}e^{-iH_{s}\left(t_{\mathrm{max}}-s_{i}^{+}\right)}\left|\beta\right\rangle .\label{eq:bare_restricted_prop_with_obs_def}
\end{equation}
These bare propagators are designated by thin solid lines in the diagrammatic
representation (see Fig.~\ref{fig:hybridization_diagram_4th}a).
The \emph{full} restricted propagator from $s_{i}$ to $s_{f}$ can
be defined in terms of Eqs.~(\ref{eq:bare_restricted_prop_def})
and (\ref{eq:bare_restricted_prop_with_obs_def}) with $H_{S}$ replaced
by $H$, and is evaluated (by way of the Dyson series) as an integral
over configurations
\begin{equation}
G_{\alpha\beta}\left(s_{f},s_{i}\right)=\int_{\boldsymbol{s}\in\left[s_{i},s_{f}\right]}\mathrm{d}\boldsymbol{s}{\cal G}_{\alpha\beta}\left(\boldsymbol{s}\right).
\end{equation}
The notation $\boldsymbol{s}\in\left[s_{i},s_{f}\right]$ indicates
that the vertex times appearing in the configuration $\boldsymbol{s}$
are restricted to the interval $\left[s_{i},s_{f}\right]$. The influence
functional then takes the same general form as Eq.~(\ref{eq:hyb_influence_functional_2})
\begin{equation}
{\cal G}_{\alpha\beta}\left(\boldsymbol{s}\right)=\left(-1\right)^{n}i^{m}{\cal U}_{\alpha\beta}^{\prime}\left(\boldsymbol{s}\right){\cal L}\left(\boldsymbol{s}\right),
\end{equation}
namely it is composed of system and bath parts, ${\cal U}_{\alpha\beta}^{\prime}\left(\boldsymbol{s}\right)$
and ${\cal L}\left(\boldsymbol{s}\right)$. The bath influence functional
is identical to the one given by Eqs.~(\ref{eq:bath_functional_sum})
and (\ref{eq:bath_functional_per_term}), and the system influence
functional will be discussed immediately below.

The system influence functional, like the bare propagator, takes
on different forms for intervals on a single branch as compared to
across branches. For a single branch interval, it is 
\begin{equation}
\begin{split} & {\cal U}_{\alpha\beta}^{\prime}\left(\boldsymbol{s}\in\left[s_{i}^{\pm},s_{f}^{\pm}\right]\right)=\\
 & \ \left\langle \alpha\right|e^{-iH_{s}s_{f}^{\pm}}\widetilde{\sigma}_{z}\left(s_{n}^{\pm}\right)\cdots\widetilde{\sigma}_{z}\left(s_{1}^{\pm}\right)e^{iH_{s}s_{i}^{\pm}}\left|\beta\right\rangle ,
\end{split}
\end{equation}
while for a cross-branch interval it becomes
\begin{equation}
\begin{split} & {\cal U}_{\alpha\beta}^{\prime}\left(\boldsymbol{s}\in\left[s_{i}^{+},s_{f}^{-}\right]\right)=\\
 & \ \left\langle \alpha\right|e^{-iH_{s}s_{f}^{-}}\widetilde{\sigma}_{z}\left(s_{m}^{-}\right)\cdots\widetilde{\sigma}_{z}\left(s_{n+1}^{-}\right)\times\\
 & ~\ \widetilde{\sigma}_{z}\left(t\right)\widetilde{\sigma}_{z}\left(s_{n}^{+}\right)\cdots\widetilde{\sigma}_{z}\left(s_{1}^{+}\right)e^{iH_{s}s_{i}^{+}}\left|\beta\right\rangle .
\end{split}
\end{equation}
Note that if $s_{i}^{+}=s_{f}^{+}$ or $s_{i}^{-}=s_{f}^{-}$, with
both times on the same branch, the restricted propagator is trivially
equal to $G^{\left(0\right)}\left(s_{f},s_{f}\right)=G\left(s_{f},s_{f}\right)=1$.
However, if $s_{i}^{+}=s_{f}^{-}$, with the times appearing on opposite
branches, $G_{\alpha\beta}\left(s_{f}^{-},s_{i}^{+}\right)$ becomes
the expectation value of the observable given that the system density
matrix was initially in the state $\left|\beta\right\rangle \left\langle \alpha\right|$:
\begin{equation}
G_{\alpha\beta}\left(s_{f}^{-},s_{f}^{+}\right)=\left\langle \left\langle \alpha\right|\sigma_{z}\left(t-s_{f}\right)\left|\beta\right\rangle \right\rangle _{b}.
\end{equation}
In terms of diagrams, the full restricted propagator is represented
by a thick segment (see Fig.~\ref{fig:hybridization_diagram_4th}a).

\subsection{Inchworm algorithm\label{subsec:coupling-inchworm}}

Suppose that the full set of restricted propagators $G_{\alpha\beta}\left(s_{k},s_{j}\right)$
for all $s_{i}<s_{j},s_{k}<s_{\uparrow}$ is known, and one wants
to evaluate a restricted propagator over a longer interval $\left[s_{i},s_{f}\right]$,
with $s_{f}>s_{\uparrow}$. It is possible to define an \emph{extended}
propagator for the interval $\left[s_{i},s_{f}\right]$ by appending
the bare propagator to the full propagator:
\begin{equation}
\overline{G}\left(s_{k},s_{j}\right)=\begin{cases}
G^{\left(0\right)}\left(s_{k},s_{j}\right) & s_{j},s_{k}>s_{\uparrow},\\
G\left(s_{k},s_{j}\right) & s_{j},s_{k}<s_{\uparrow},\\
G^{\left(0\right)}\left(s_{k},s_{\uparrow}\right)G\left(s_{\uparrow},s_{j}\right) & s_{j}<s_{\uparrow}<s_{k}.
\end{cases}
\end{equation}
Since the contributions of all configurations $\boldsymbol{s}\in\left[s_{i},s_{\uparrow}\right]$
are included in the extended propagator, it is in fact only necessary
to sum over configurations in which at least one vertex is contained
in the interval $\left[s_{\uparrow},s_{f}\right]$. The propagator
over the entire interval $\left[s_{i},s_{f}\right]$ can be constructed
as a path integral over configurations
\begin{equation}
G\left(s_{f},s_{i}\right)=\int_{\boldsymbol{s}\in\left[s_{i},s_{f}\right]}\mathrm{d}\boldsymbol{s}\overline{{\cal G}}_{\alpha\beta}\left(\boldsymbol{s}\right).
\end{equation}
The influence functional $\overline{{\cal G}}_{\alpha\beta}$ is defined
in terms of extended propagators and a modified bath influence functional.
It takes the form
\begin{equation}
\overline{{\cal G}}\left(\boldsymbol{s}\right)=\overline{G}\left(s_{f},s_{m}\right)\cdots\overline{G}\left(s_{2},s_{1}\right)\overline{G}\left(s_{1},s_{i}\right)\times\sum_{q\in{\cal Q^{\prime}}_{m}}{\cal L}_{q}\left(\boldsymbol{s}\right).
\end{equation}
The bath influence functional $\sum_{q\in{\cal Q^{\prime}}_{m}}{\cal L}_{q}\left(\boldsymbol{s}\right)$
is similar to that of Eq.~(\ref{eq:hybridization_functional}), but
summation is only carried out over ${\cal Q}_{m}^{\prime}\subseteq{\cal Q}_{m}$,
a subset of the pairings including only \emph{inchworm proper} pairings.

\begin{figure}
\begin{raggedright}
(a)
\par\end{raggedright}
\begin{centering}
\includegraphics{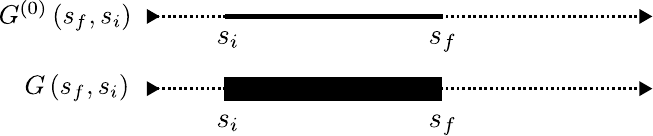}
\par\end{centering}
\begin{raggedright}
(b)
\par\end{raggedright}
\begin{centering}
\includegraphics{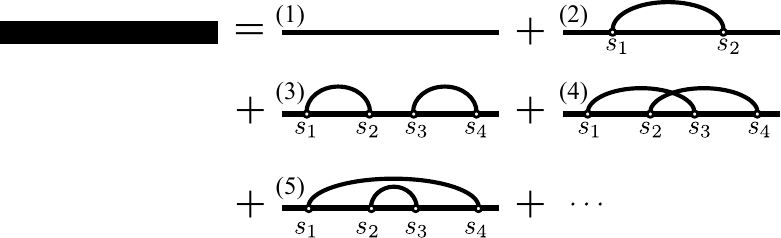}
\par\end{centering}
\begin{raggedright}
(c)
\par\end{raggedright}
\centering{}\includegraphics{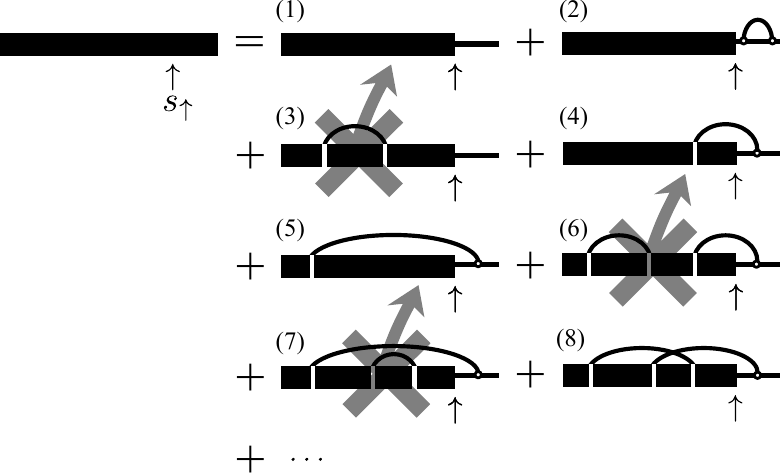}\caption{(a) Diagrammatic representation of the bare restricted propagator
$G^{\left(0\right)}$ (thin solid line) and the full restricted propagator
$G$ (thick solid line) of the subinterval $\left[s_{i},s_{f}\right]$
on an unfolded Keldysh contour. (b) The bare dQMC expression for the
system\textendash bath coupling expansion. The arched curves connecting
pairs of vertices within each configuration describe the coupling
interaction. Diagram (b.1) is the zeroth order contribution. Diagram
(b.2) is the diagram associated with a given $2^{\mathrm{nd}}$ order
configuration $(s_{1},s_{2})$. Diagrams (b.3)\textendash (b.5) are
three diagrams (corresponding to three possible pairings of ${\cal Q}_{4}$)
associated with a $4^{\mathrm{th}}$ order configuration $(s_{1},s_{2},s_{3},s_{4})$.
(c) The inchworm algorithm in the system\textendash bath coupling
expansion. All the full restricted propagators are assumed to be known
for any subinterval to the left of the $s_{\uparrow}$ time. Diagram
(c.1) is the zeroth order inchworm diagram. Diagram (c.2), (c.4) and
(c.5) are all inchworm proper $2^{\mathrm{nd}}$ order diagrams. Diagrams
(c.3) is an inchworm improper diagram that is included in diagram
(c.1). Diagrams (c.6)\textendash (c.8) are associated with the same
$4^{\mathrm{th}}$order configuration. Diagrams (c.6) and (c.7) are
included in diagrams (c.4) and (c.5), respectively and only diagram
(c.8) is inchworm proper. \label{fig:hybridization_diagram_4th}}
\end{figure}

To define inchworm propriety, we first define two pairs to be \emph{connected}
if their interaction lines, which are drawn between the members of
each pair, cross each other. As connectedness is clearly an equivalence
relation, any pairing can be partitioned into disjoint sets of connected
pairs, called ``clusters''. A pairing or diagram is inchworm proper
if this procedure does not generate a cluster with all of its vertices
contained in $\left[s_{i},s_{\uparrow}\right]$. Put differently,
to check whether a particular diagram is inchworm proper one should
cluster together sets of interaction lines which cross each other.
If and only if every cluster includes at least one line with an endpoint
in $\left[s_{\uparrow},s_{f}\right]$ is the diagram inchworm proper.
This is illustrated in Fig.~\ref{fig:hybridization_diagram_4th},
where two examples of improper diagrams are crossed out.

It is straightforward to prove that any diagram in the bare expansion
is accounted for once and only once within the inchworm scheme, and
that an inchworm diagram does not generate any spurious diagrams not
included in the bare scheme; therefore, it is formally exact. \emph{However,
every inchworm diagram contains an infinite number of bare diagrams,
making the expansion substantially more efficient than the bare one.}

This method will be referred to as the System\textendash Bath Coupling
Inchworm (SBCI) approach in the companion paper.

\section{Diabatic Coupling Expansion\label{sec:Diabatic-Coupling-Expansion}}

\subsection{Polaron transformation}

We now consider an expansion in terms of the diabatic coupling $H^{\prime}=\Delta\sigma_{x}$,
i.e. the spin-flip interaction. Analogous approaches for the Anderson\textendash Holstein
model have been very successful in certain regimes.\cite{Muehlbacher2008Realtimepath,Werner2007Efficientdynamicalmean,Werner2013Phononenhancedrelaxation,chen_anderson-holstein_2016}
The unperturbed Hamiltonian is in this case $H_{0}=H_{b}+\sigma_{z}\left(\epsilon+\sum_{k}c_{k}x_{k}\right)$.
Since $H_{0}$ commutes with $\sigma_{z}$, its eigenstates maintain
the spin quantum number $\sigma=\pm1$, which partitions them into
two subspaces. Within each subspace the Hamiltonian is easily diagonalized
by a polaron transformation. The effective Hamiltonian for the $\sigma=+1$
and $\sigma=-1$ subspaces, respectively, is
\begin{equation}
H_{\sigma}=H_{b}+\sigma\left(\epsilon+\sum_{\ell}c_{\ell}x_{\ell}\right).
\end{equation}
We apply the (canonical) transformation
\begin{equation}
{\cal B}_{\sigma}H_{\sigma}{\cal B}_{\sigma}^{\dagger}=H_{b}+\sigma\epsilon-\sum_{\ell}\frac{c_{\ell}^{2}}{\omega_{\ell}^{2}},
\end{equation}
where

\begin{align}
{\cal B}_{\sigma}= & e^{\theta_{\sigma}},\\
\theta_{\sigma}= & \sigma\sum_{\ell}\frac{c_{\ell}}{\omega_{\ell}^{3/2}}\left(b_{\ell}^{\dagger}-b_{\ell}\right).
\end{align}
Since $\theta_{+}=-\theta_{-}$, it is convenient to write ${\cal B}_{\sigma}^{\dagger}={\cal B}_{\bar{\sigma}}$.
We also define $\epsilon_{\sigma}=\sigma\epsilon-\sum_{\ell}\frac{c_{\ell}^{2}}{\omega_{\ell}^{2}}$.
With these definitions, the unperturbed propagator can be written
in the form 
\begin{equation}
e^{-iH_{0}t}=\sum_{\sigma=\pm}e^{-i\epsilon_{\sigma}t}{\cal B}_{\bar{\sigma}}e^{-iH_{b}t}{\cal B}_{\sigma}\left|\sigma\right\rangle \left\langle \sigma\right|.
\end{equation}
In this form the interaction picture time evolution will turn out
to be very easy to evaluate, as discussed below.

The natural initial condition for the expansion in the diabatic coupling
is $\rho_{b}=\exp\left[-\beta H_{\pm}\right]$, and using one of these
two choices simplifies the expressions substantially. However, in
order to allow for rigorous comparison with the system\textendash bath
coupling expansion, we choose to start from a state described by $\rho_{b}=\exp\left[-\beta H_{b}\right]$.
Unfortunately, this introduces additional complications in the expressions
given below, and we will comment on this as we proceed. The choice
of initial condition does not otherwise impact the formalism.

\subsection{Bare dQMC }

To obtain a dQMC algorithm for the expectation value of $O=\hat{\sigma}_{z}$,
we must write the contribution ${\cal O}\left(\boldsymbol{s}\right)$
of a configuration $\mathbf{s}$ in the form of Eq.~\ref{eq:functional_general}.
In the interaction picture, $\widetilde{\sigma}_{x}\left(s\right)=e^{iH_{0}s}\sigma_{x}e^{-iH_{0}s}$,
we can write
\begin{equation}
\begin{split}{\cal O}\left(\boldsymbol{s}\right)=\left(-1\right)^{n}i^{m}\Delta^{m} & \Bigl\langle\widetilde{\sigma}_{x}\left(s_{m}\right)\cdots\widetilde{\sigma}_{x}\left(s_{n+1}\right)\times\\
 & \widetilde{\sigma}_{z}\left(t\right)\widetilde{\sigma}_{x}\left(s_{n}\right)\cdots\widetilde{\sigma}_{x}\left(s_{1}\right)\Bigr\rangle.
\end{split}
\label{eq:functional_int}
\end{equation}
We designate the state between $\left[s_{k,}s_{k+1}\right]$ as $\sigma_{k+1}$
for $k\in\left\{ 0,\ldots,m-1\right\} $, with $s_{0}\equiv0^{+}$
and $s_{m+1}\equiv0^{-}$. The observable $\sigma_{z}$ at the tip
of the contour does not change the state, while every application
of $\sigma_{x}$ flips the state from $\sigma$ to $\bar{\sigma}$.
Since the initial condition of the spin is specified $\rho_{s}=\left|1\right\rangle \left\langle 1\right|=\left|+\right\rangle \left\langle +\right|$,
we have $\sigma_{1}=\sigma_{m+1}=+$. The contribution ${\cal O}\left(\boldsymbol{s}\right)$
of a configuration $\boldsymbol{s}$ to the expectation value of $O=\hat{\sigma}_{z}$
can then be expressed as a product of a system influence functional
$\Phi\left(\boldsymbol{s}\right)$ and a bath influence functional
${\cal J}\left(\boldsymbol{s}\right)$:
\begin{equation}
{\cal O}\left(\boldsymbol{s}\right)=\left(-1\right)^{n}i^{m}\Delta^{m}\Phi\left(\boldsymbol{s}\right){\cal J}\left(\boldsymbol{s}\right).
\end{equation}
The system functional $\Phi\left(\boldsymbol{s}\right)$ handles the
influence of propagation within the system,
\begin{equation}
\begin{split}\Phi\left(\boldsymbol{s}\right)= & \left\langle +1\right|\sigma_{x}^{n'}\sigma_{z}\sigma_{x}^{n}\left|+1\right\rangle \\
 & \ \times\exp\left[-i\sum_{k=1}^{m+1}\epsilon_{\sigma_{k}}\left(s_{k}-s_{k-1}\right)\right],
\end{split}
\end{equation}
whiles the bath functional ${\cal J}\left(\boldsymbol{s}\right)$
is a multi-time correlation function of bath operators
\begin{equation}
{\cal J}\left(\boldsymbol{s}\right)=\Bigl\langle\widetilde{{\cal B}}_{-}\left(s_{m+1}\right)\prod_{k=1}^{m}\widetilde{{\cal B}}_{\bar{\sigma}_{k}}^{2}\left(s_{k}\right)\widetilde{{\cal B}}_{+}\left(s_{0}\right)\Bigr\rangle_{b}.\label{eq:multi-time-polaron-correlation}
\end{equation}
Here $\left\langle \cdot\right\rangle _{b}=\mathrm{Tr}_{b}\left\{ \rho_{b}\cdot\right\} $
and $\rho_{b}$ is the initial bath density matrix. The first and
last factors are induced by the initial condition. By a generalized
Wick's theorem for polaron shift operator (see Appendix~\ref{sec:Wick's-Theorem-for}),
we can write $\mathcal{J}\left(\mathbf{s}\right)$ as a product of
two\textendash time correlation functions, 
\begin{equation}
{\cal J}\left(\boldsymbol{s}\right)=\frac{\prod_{\left(j,k\right)\in{\cal C}_{m+2}^{\mathrm{odd}}}C\left(s_{k},s_{j}\right)^{r_{k}r_{j}}}{\prod_{\left(j,k\right)\in{\cal C}_{m+2}^{\mathrm{even}}}C\left(s_{k},s_{j}\right)^{r_{k}r_{j}}}.
\end{equation}
where $r_{i}=1$ if $i=1,m+1$, otherwise $r_{i}=2$. The fact that
the powers in the numerator and denominator may differ arises from
the initial condition. Here we have defined 
\begin{equation}
{\cal C}_{m+2}^{\mathrm{even}}=\left\{ \left.\left(j,k\right)\in{\cal C}_{m+2}\right|\left|k-j\right|~\mathrm{even}\right\} ,
\end{equation}
and 
\begin{equation}
{\cal C}_{m+2}^{\mathrm{odd}}=\left\{ \left.\left(j,k\right)\in{\cal C}_{m+2}\right|\left|k-j\right|~\mathrm{odd}\right\} .
\end{equation}
which are subsets of all possible pairings of $m+2$ elements. The
pairings of $m$ elements, $\mathcal{C}_{m}$, denotes the set of
all ordered tuples composed of different integers between 0 and $m-1$.
For example,
\[
{\cal C}_{2}^{\mathrm{odd}}=\left\{ \left(0,1\right)\right\} ,\ {\cal C}_{2}^{\mathrm{even}}=\left\{ \right\} ,
\]
where $\left\{ \right\} $ denotes the empty set and
\[
\begin{split} & {\cal C}_{4}^{\mathrm{odd}}=\left\{ \left(0,1\right),\left(1,2\right),\left(2,3\right),\left(0,3\right)\right\} ,\\
 & {\cal C}_{4}^{\mathrm{even}}=\left\{ \left(0,2\right),\left(1,3\right)\right\} .
\end{split}
\]
The correlation function $C\left(s_{k},s_{j}\right)$ is one of the
expressions complicated by the initial condition, and is given by
\begin{equation}
C\left(s_{k},s_{j}\right)=\left\langle \widetilde{{\cal B}}_{-}\left(s_{k}\right)\widetilde{{\cal B}}_{+}\left(s_{j}\right)\right\rangle _{b}\label{eq:correlation_function_inBB}
\end{equation}
In general, we can write the two-time correlation function of the
polaron shift operator as (see Appendix.~(\ref{sec:Wick's-Theorem-for}))
\begin{eqnarray}
C\left(s_{k},s_{j}\right) & = & e^{-Q_{2}\left(s\right)-iQ_{1}\left(s\right)},\label{eq:correlation_function_inQ1Q2}
\end{eqnarray}
with
\begin{equation}
Q_{1}\left(s\right)=\frac{2}{\pi}\int\mathrm{d}\omega\frac{J\left(\omega\right)}{\omega^{2}}\sin\omega s,\label{eq:Q1}
\end{equation}
and
\begin{equation}
Q_{2}\left(s\right)=\frac{2}{\pi}\int\mathrm{d}\omega\frac{J\left(\omega\right)}{\omega^{2}}\coth\left(\frac{\beta\omega}{2}\right)\left(1-\cos\omega s\right).\label{eq:Q2}
\end{equation}
In the diagrammatic representation shown in Fig.~\ref{fig:interaction_diagram},
the two-time correlation function is represented by dashed lines.
There exists an extra set of lines due to the initial condition, which
connect every vertex to the edges of the diagram. To avoid overcrowding
the diagram with information, these are not shown. A dashed line above
the contour describes a contribution to the numerator, while one under
the contour describes one associated with the denominator. Each vertex
is connected by such interaction lines to every other vertex in the
configuration, and since only one way to do this exists, each configuration
generates exactly one diagram. The bare Monte Carlo implementation
based on this expansion is illustrated in Fig.~\ref{fig:interaction_diagram}b.

\subsection{Inchworm algorithm}

\begin{figure}
\begin{raggedright}
(a)
\par\end{raggedright}
\begin{centering}
\includegraphics{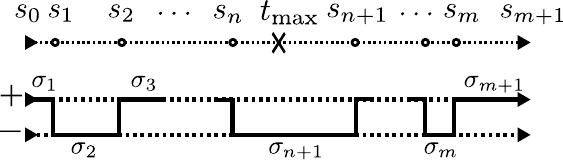}
\par\end{centering}
\begin{raggedright}
(b)
\par\end{raggedright}
\begin{centering}
\includegraphics{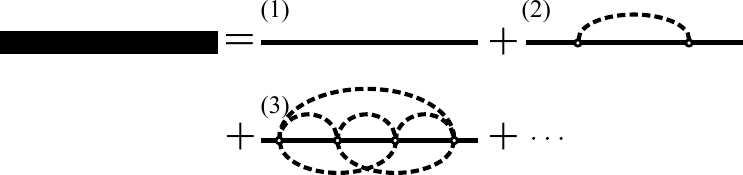}
\par\end{centering}
\begin{raggedright}
(c)
\par\end{raggedright}
\begin{centering}
\includegraphics{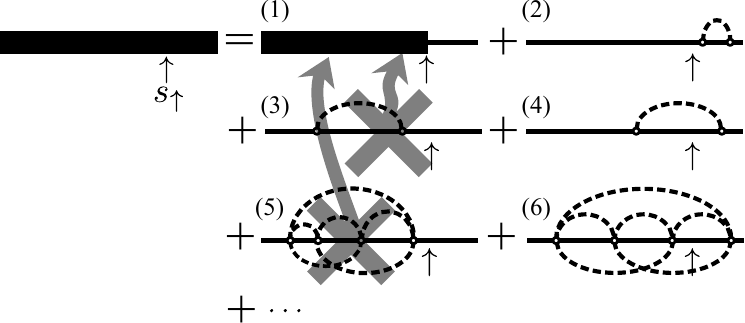}
\par\end{centering}
\caption{(a) A configuration including $s_{0}=s_{i}$ and $s_{m+1}=s_{f}$
for the diabatic coupling expansion. The state of the system flips
at every $s_{i}$. (b) Diagrams appearing in bare dQMC. The dashed
curve $\left(12\right)$ indicates an interaction line in either the
numerator (above the contour) or the denominator (below it). Only
one diagram corresponds to each configuration. (c) The naive inchworm
scheme. Diagrams with no vertices after $s_{\uparrow}$ (such as (c.3)
and (c.5)) are contained in the zeroth order term (c.1) and need not
be summed. Other diagrams, such as (c.4) and (c.6), are analogous
to those of the bare dQMC.\label{fig:interaction_diagram}}
\end{figure}

The process of formulating an inchworm expansion is analogous to
that of Sec.~\ref{subsec:coupling-inchworm}, but with the diagrammatic
structure of the diabatic coupling expansion. Inchworm proper and
improper diagrams are illustrated in Fig.~\ref{fig:interaction_diagram}c.
The main difference is that whereas diagrams in the system\textendash bath
coupling expansion include interaction lines only between vertices
paired within a particular pairing, the diabatic coupling expansion
includes an interaction line between every two vertices. Therefore,
there is only one ``cluster'' of vertices in every diagram, and
that diagram is required to have at least one vertex in $\left[t_{\uparrow},t_{f}\right]$.
The only diagram not containing such a cluster is the order zero diagram
(shown as (1) in Fig.~\ref{fig:interaction_diagram}c). This is also
the only diagram containing an infinite number of bare diagrams: each
diagram containing a cluster is completely identical to the one and
only bare diagram that it represents.

The main advantages of the inchworm algorithm are therefore lost
in the direct diabatic coupling scheme described here, and indeed
we have verified that upon implementation of such an algorithm an
exponential dynamical sign problem appears (not shown). However, in
the remainder of this paper, this problem is circumvented by transforming
the expansion to a cumulant form. From this perspective, a very useful
inchworm algorithm then emerges.

\section{Diabatic Coupling Cumulant Inchworm (DCCI) Expansion\label{sec:Diabatic-Coupling-Cumulant}}

As noted in Sec.~\ref{sec:Diabatic-Coupling-Expansion}, the diabatic
coupling expansion in its direct Keldysh formulation has a peculiar
diagrammatic structure in which each interaction vertex is directly
connected to every other vertex. As such, this expansion does not
significantly benefit from the inchworm algorithm, which relies on
the ability to cut diagrams into weakly-connected subgraphs. We now
show that by reformulating the diabatic coupling expansion in cumulant
form, one obtains a formalism which is much more amenable to inchworm
dQMC. The cumulant formalism has the additional advantage of being
written in physical (rather than contour) time, such that in the absence
of a sign problem the computation scales linearly with time, as will
be demonstrated in the companion paper.

Since cumulants are most conveniently defined in terms of moments,
the moment form of the expansion will first be presented. Cumulants
and the cumulant inchworm algorithm will then be presented.

\subsection{Moments}

Consider the evaluation of the dynamics of an observable $O$ in
terms of its \emph{moments}, $\mu_{m}\left(\tau_{1},...,\tau_{m}\right)$.
Given that we have Eq.~(\ref{eq:path_integral_1}), such that $\left\langle O\left(t\right)\right\rangle =\int\mathrm{d}\boldsymbol{s}{\cal O}\left(\boldsymbol{s}\right)$,
the observable can be written in terms of a moment expansion
\begin{equation}
\int\mathrm{d}\boldsymbol{s}{\cal O}\left(\boldsymbol{s}\right)=\sum_{m=0}^{\infty}\int\mathrm{d}\boldsymbol{\tau}\mu_{m}\left(\tau_{1},\tau_{2},\cdots,\tau_{m}\right).\label{eq:moment_expansion}
\end{equation}
While the integration $\int\mathrm{d}\boldsymbol{s}$ is performed
over contour time, the integration $\int\mathrm{d}\boldsymbol{\tau}$
is performed over physical times $\tau_{1}>\tau_{2}>\cdots>\tau_{m}$,
such that

\begin{equation}
\int\mathrm{d}\boldsymbol{\tau}=\int_{0}^{t}\mathrm{d}\tau_{1}\int_{0}^{\tau_{1}}\mathrm{d}\tau_{2}\cdots\int_{0}^{\tau_{m-1}}\mathrm{d}\tau_{m}.
\end{equation}
An $m^{\mathrm{th}}$-order moment $\mu_{m}\left(\tau_{1},\tau_{2},\cdots,\tau_{m}\right)$
is defined as
\begin{equation}
\mu_{m}\left(\tau_{1},...,\tau_{m}\right)=\sum_{\alpha_{i}\in\left\{ +,-\right\} }{\cal O}\left({\cal T}_{c}\left[\tau_{1}^{\alpha_{1}},...,\tau_{m}^{\alpha_{m}}\right]\right),\label{eq:moment_definition}
\end{equation}
where ${\cal T}_{c}$ indicates contour time ordering and the $\alpha_{i}=\pm$
are the Keldysh branch indices. The moments are defined as functions
of a set of real times, and the Keldysh branch indices are summed
over. This is equivalent to simultaneously collecting the contributions
from entire classes of path configurations associated with the real
times, $\tau_{1},...,\tau_{m}$, as illustrated diagrammatically in
Fig.~\ref{fig:moment_and_cumulant}a. Notably, it is exponentially
expensive as a function of the order $m$ to evaluate moments in terms
of diagrams, as an $m^{\mathrm{th}}$ order moment is the sum of $2^{m}$
diagrams.

For the population operator $O=\sigma_{z}$ in the diabatic coupling
expansion, the $0^{\mathrm{th}}$ order moment is $\mu_{0}=1$ and
odd moments vanish, $\mu_{2n+1}=0$. The expectation value of $\sigma_{z}$
can therefore be written in terms of the even moments
\begin{equation}
\begin{split}\left\langle \sigma_{z}\left(t\right)\right\rangle ={} & 1+\int_{0}^{t}\mathrm{d}\tau_{1}\int_{0}^{\tau_{1}}\mathrm{d}\tau_{2}\mu_{2}\left(\tau_{1},\tau_{2}\right)\\
 & +\int_{0}^{t}\mathrm{d}\tau_{1}\int_{0}^{\tau_{1}}\mathrm{d}\tau_{2}\int_{0}^{\tau_{2}}\mathrm{d}\tau_{3}\int_{0}^{\tau_{3}}\mathrm{d}\tau_{4}\\
 & ~\times\mu_{4}\left(\tau_{1},\tau_{2},\tau_{3},\tau_{4}\right)\\
 & +\ldots\ .
\end{split}
\label{eq:moment_expansion_4th}
\end{equation}
With the initial density matrix $\left|1\right\rangle \left\langle 1\right|e^{-\beta H_{b}}$
specified earlier, the second population moment simplifies to\cite{Reichman1997Cumulantexpansionsand}
\begin{equation}
\mu_{2}\left(\tau_{1},\tau_{2}\right)=-4\Delta^{2}\text{Re}\left\{ e^{2i\epsilon\left(\tau_{1}-\tau_{2}\right)}{\cal J}\left(0^{+},\tau_{2}^{+},\tau_{1}^{+},0^{-}\right)\right\} ,
\end{equation}
and the fourth moment to
\begin{equation}
\begin{split}\mu_{4} & \left(\tau_{1},\tau_{2},\tau_{3},\tau_{4}\right)=4\Delta^{2}\times\text{Re}\biggl\{\\
 & e^{2i\epsilon\left(\tau_{1}-\tau_{2}+\tau_{3}-\tau_{4}\right)}{\cal J}\left(0^{+},\tau_{4}^{+},\tau_{3}^{+},\tau_{2}^{+},\tau_{1}^{+},0^{-}\right)+\\
 & e^{2i\epsilon\left(\tau_{1}-\tau_{2}-\tau_{3}+\tau_{4}\right)}{\cal J}\left(0^{+},\tau_{4}^{+},\tau_{3}^{+},\tau_{1}^{-},\tau_{2}^{-},0^{-}\right)+\\
 & e^{-2i\epsilon\left(\tau_{1}-\tau_{2}-\tau_{3}+\tau_{4}\right)}{\cal J}\left(0^{+},\tau_{4}^{+},\tau_{2}^{+},\tau_{1}^{-},\tau_{3}^{-},0^{-}\right)+\\
 & e^{-2i\epsilon\left(\tau_{1}-\tau_{2}+\tau_{3}-\tau_{4}\right)}{\cal J}\left(0^{+},\tau_{3}^{+},\tau_{2}^{+},\tau_{1}^{+},\tau_{4}^{-},0^{-}\right)\biggr\}.
\end{split}
\end{equation}

Evaluating the moments within dQMC is therefore an alternative scheme
for calculating dynamics. While linear in time (rather than quadratic,
like the Keldysh formalism which involves two times), this expansion
involves an additional exponential cost in the diagram order, due
to the summation over the Keldysh indices. However, bare moment expansions
typically converge very slowly if at all, and hold no real advantage
over a direct calculation (though they may be of help with sign problems
in certain cases\cite{Profumo2015QuantumMonteCarlo}). It is therefore
often advantageous to resum moments into cumulants. It turns out that
a relationship exists between cumulant resummation and the inchworm
algorithm, and this will be shown below.

\subsection{Cumulants\label{subsec:Cumulants}}

\begin{figure}
\begin{raggedright}
(a)
\par\end{raggedright}
\begin{centering}
\includegraphics{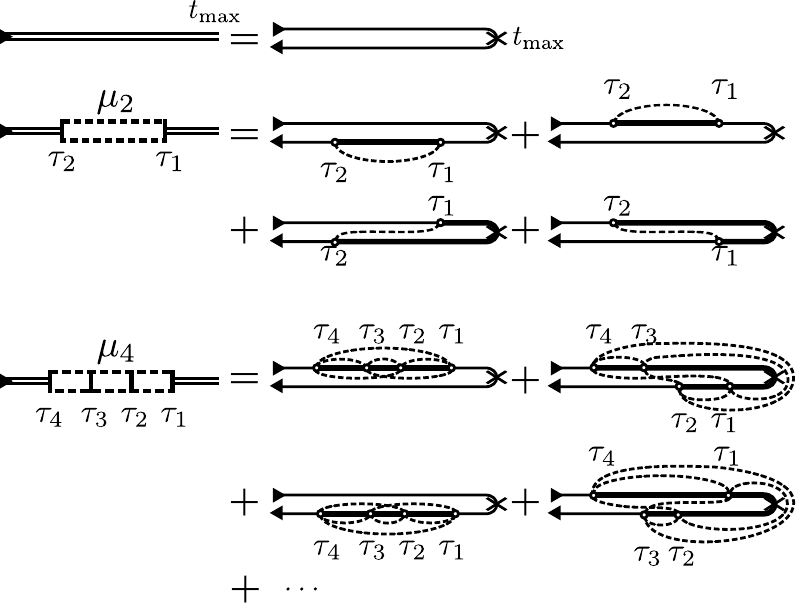}
\par\end{centering}
\begin{raggedright}
(b)
\par\end{raggedright}
\begin{centering}
\includegraphics{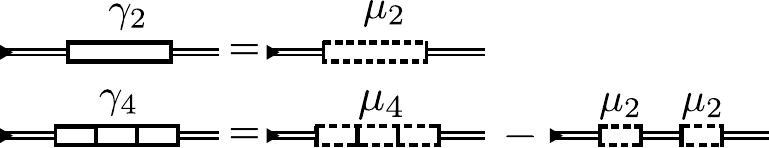}
\par\end{centering}
\caption{(a) The real-time coordinate is represented by the thin double lines.
The bare double line segment $\left[0,t_{\mathrm{max}}\right]$ corresponds
to the bare propagator in the diabatic expansion on the Keldysh contour
folded at $t_{\mathrm{max}}$. A $m^{\mathrm{th}}$ order moment of
a real time configuration $\left(\tau_{1},\tau_{2}\cdots,\tau_{m}\right)$
is illustrated as a dashed-edged box from $\tau_{1}$ to $\tau_{m}$
with solid vertical ticks at each configuration time. There are 4
distinct diagrams on the Keldysh contour associated with the $2^{\mathrm{nd}}$
moment $\mu_{2}\left(\tau_{1},\tau_{2}\right)$: $\boldsymbol{s}=\left(\tau_{1}^{-},\tau_{2}^{-}\right),\left(\tau_{2}^{+},\tau_{1}^{+}\right),\left(\tau_{1}^{+},\tau_{2}^{-}\right),\left(\tau_{2}^{+},\tau_{1}^{-}\right)$.
These diagrams are plotted by connecting the vertices with the diabatic
interaction lines as in Fig.~(\ref{fig:interaction_diagram}). The
$4^{\mathrm{th}}$ moment contains $2^{4}$ diagrams on the contour.
Here, we demonstrate only 4 example diagrams. (b) The COP cumulants
of a real-time configuration $\left(\tau_{1},\tau_{2}\cdots,\tau_{m}\right)$
are illustrated as a solid-edged box with vertical ticks at each configuration
time. Here, we show the diagrammatic representation of Eq.~(\ref{eq:cumulant_in_moment_2})
and (\ref{eq:cumulant_in_moment_4}), which illustrate the $2^{\mathrm{nd}}$
and $4^{\mathrm{th}}$ cumulants in terms of the moments.\label{fig:moment_and_cumulant}}
\end{figure}

Moment expansions can be immediately resummed into cumulant expansions
in several ways.\cite{Reichman1997Cumulantexpansionsand,VanKampen1974cumulantexpansionstochastic}
For the present purpose, it is advantageous to choose the chronological
ordering prescription (COP) cumulant expansion,\cite{Reichman1997Cumulantexpansionsand}
which yields the time-nonlocal\emph{ }equation of motion

\begin{equation}
\begin{split}\frac{\mathrm{d}\left\langle \sigma_{z}\left(t\right)\right\rangle }{\mathrm{d}t}={} & \sum_{m=2}^{\infty}\int_{0}^{t}\mathrm{d}\tau_{1}\int_{0}^{\tau_{1}}\mathrm{d}\tau_{2}...\int_{0}^{\tau_{m-2}}\mathrm{d}\tau_{m-1}\\
 & \times\gamma_{m}\left(t,\tau_{1},...,\tau_{m-1}\right)\left\langle \sigma_{z}\left(\tau_{m-1}\right)\right\rangle .
\end{split}
\label{eq:cumulant_expansion_differential}
\end{equation}
An advantage from the inchworm perspective is immediately apparent:
the expression depends on the population at shorter times, such that
previously calculated properties can perhaps be reused. The $m$-th
order COP cumulant $\gamma_{m}\left(t,\tau_{1},...,\tau_{m-1}\right)$
can be obtained by plugging Eq.~(\ref{eq:moment_expansion_4th})
into both sides of Eq.~(\ref{eq:cumulant_expansion_differential})
and equating terms of equal order. For example,
\begin{equation}
\gamma_{2}\left(\tau_{1},\tau_{2}\right)=\mu_{2}\left(\tau_{1},\tau_{2}\right),\label{eq:cumulant_in_moment_2}
\end{equation}
\begin{equation}
\gamma_{4}\left(\tau_{1},\tau_{2},\tau_{3},\tau_{4}\right)=\mu_{4}\left(\tau_{1},\tau_{2},\tau_{3},\tau_{4}\right)-\mu_{2}\left(\tau_{1},\tau_{2}\right)\mu_{2}\left(\tau_{3},\tau_{4}\right),\label{eq:cumulant_in_moment_4}
\end{equation}
and $\gamma_{2n-1}=0$. A general $m$-th order cumulant, $\gamma_{m}$,
can be obtained recursively: 
\begin{equation}
\begin{split}\gamma_{m}\left(\tau_{1},...,\tau_{m}\right) & =\sum_{p\in{\cal P}_{m}}-\left(-1\right)^{\left|p\right|}\times\\
 & \prod_{\left(i_{1},i_{2},\cdots,i_{2n}\right)\in p}\mu_{2n}\left(\tau_{i_{1}},\tau_{i_{2}}\cdots,\tau_{i_{2n}}\right).
\end{split}
\end{equation}
The set ${\cal P}_{m}$ describes all possible ways of partitioning
a sequence of integers $1,2,\dots,m$ into subsequences of adjacent
numbers, each having an even number elements. Each partition $p\in\mathcal{P}_{m}$
can be represented by a set of ordered tuples $\left(i_{1},i_{2},\dots,i_{2n}\right)$
corresponding to one subsequence, and $\left|p\right|$ is the number
of subsequences within the partition. For instance,
\[
{\cal P}_{2}=\left\{ \left\{ \left(1,2\right)\right\} \right\} ,
\]
\[
{\cal P}_{4}=\left\{ \left\{ \left(1,2,3,4\right)\right\} ,\left\{ \left(1,2\right),\left(3,4\right)\right\} \right\} ,
\]
\[
\begin{split}{\cal P}_{6}= & \left\{ \left\{ \left(1,2,3,4,5,6\right)\right\} ,\left\{ \left(1,2\right),\left(3,4,5,6\right)\right\} ,\right.\\
 & \left.\ \left\{ \left(1,2,3,4\right),\left(5,6\right)\right\} ,\left\{ \left(1,2\right),\left(3,4\right),\left(5,6\right)\right\} \right\} .
\end{split}
\]
The diagrammatic description of COP cumulants in terms of moments
is shown in Fig.~\ref{fig:moment_and_cumulant}b. A cumulant of any
given order can be expressed in terms of moments up to and including
the same order.

\subsection{Naive inchworm algorithm\label{subsec:Naive-inchworm-algorithm}}

The dynamics of $\left\langle \sigma_{z}\left(t\right)\right\rangle $
within the COP cumulant expansion can be evaluated by dQMC. To simplify
the notation, it is convenient to redefine the times $t,\tau_{1},...,\tau_{m-1}$
as $\tau_{1},\tau_{2}...,\tau_{m}$, respectively; these obey the
physical time ordering $\tau_{1}>\cdots>\tau_{m}$. As before, the
times will be indicated by the vector quantity $\boldsymbol{\tau}$
when possible. By carrying out the integration $\int_{0}^{t}\mathrm{d}\tau_{1}$
on both sides of Eq.~(\ref{eq:cumulant_expansion_differential}),
an expression for $\left\langle \sigma_{z}\left(t\right)\right\rangle $
in terms of itself is obtained: 
\begin{equation}
\left\langle \sigma_{z}\left(t\right)\right\rangle =1+\int_{0}^{t}\mathrm{d}\boldsymbol{\boldsymbol{\tau}}\mathcal{K}\left(\boldsymbol{\boldsymbol{\tau}}\right),\label{eq:cumulant_expansion_integral}
\end{equation}
 
\begin{equation}
\mathcal{K}\left(\boldsymbol{\boldsymbol{\tau}}\right)=\gamma_{m}\left(\tau_{1},...,\tau_{m}\right)\left\langle \sigma_{z}\left(\tau_{m}\right)\right\rangle .\label{eq:cumulant_expansion_functional_withP}
\end{equation}
Since $\gamma_{2n-1}=0$, the path integration $\int_{0}^{t}\mathrm{d}\boldsymbol{\boldsymbol{\tau}}$
can be explicitly written as
\begin{equation}
\int_{0}^{t}\mathrm{d}\boldsymbol{\boldsymbol{\tau}}=\sum_{m\in\mathrm{even},\ge2}\int_{0}^{t}\mathrm{d}\tau_{1}\int_{0}^{\tau_{1}}\mathrm{d}\tau_{2}...\int_{0}^{\tau_{m-1}}\mathrm{d}\tau_{m}.
\end{equation}

\begin{figure}
\begin{centering}
\includegraphics{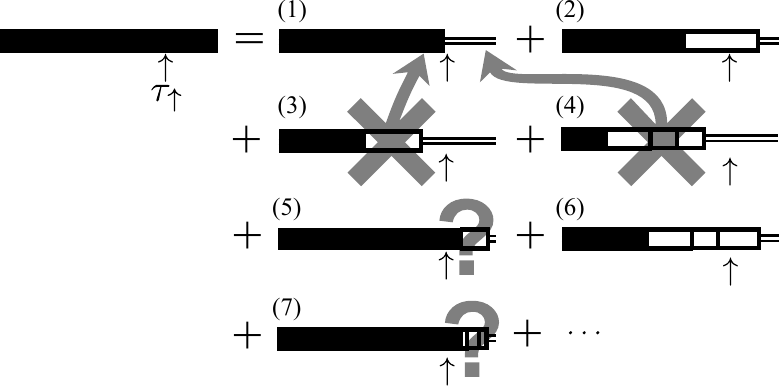}
\par\end{centering}
\caption{Diagrammatic representation of the naive prescription of the inchworm
algorithm, Eq.~(\ref{eq:cumulant_inchworm_naive}). The solid-edged
boxes with vertical ticks are the COP cumulants as shown in Fig.~(\ref{fig:moment_and_cumulant}).
The $\tau_{\uparrow}$ is indicated as the $\uparrow$ on the physical
time coordinate. Each configuration corresponds to one single diagram.
Diagrams (3) and (4) have all cumulant boxes lying in the known region
(to the left of $\uparrow$) and are considered been included in diagram
(1) for this inchworm step. The cumulant boxes in diagrams (2) and
(6) straddle the $\tau_{\uparrow}$ time and their contribution can
be calculated by Eq.~(\ref{eq:cumulant_expansion_functional_withP}).
Diagrams (5) and (7) have all cumulant boxes located to the right
of the $\uparrow$, are unknown for this inchworm step in the naive
version. \label{fig:cumulant_naive_inchworm}}
\end{figure}

Since the functional $\mathcal{K}\left(\boldsymbol{\boldsymbol{\tau}}\right)$
depends on $\left\langle \sigma_{z}\left(t_{m}\right)\right\rangle $,
the observable is evaluated at the smallest time in the configuration
$\boldsymbol{\tau}$. Since this is the quantity being evaluated,
it is not known to begin with and there is no bare expansion of the
COP type. However, it is straightforward to implement a simple inchworm
algorithm: assume $\left\langle \sigma_{z}\left(\tau\right)\right\rangle $
is known for all $\tau\in\left[0,\tau_{\uparrow}\right]$. The expectation
value at $t>\tau_{\uparrow}$ can then be expressed as:
\begin{equation}
\left\langle \sigma_{z}\left(t\right)\right\rangle =\left\langle \sigma_{z}\left(\tau_{\uparrow}\right)\right\rangle +\int_{\tau_{\uparrow}}^{t}\mathrm{d}\boldsymbol{\boldsymbol{\tau}}\mathcal{K}\left(\boldsymbol{\boldsymbol{\tau}}\right).\label{eq:cumulant_inchworm_naive}
\end{equation}
Here $\int_{\tau_{\uparrow}}^{t}\mathrm{d}\boldsymbol{\boldsymbol{\tau}}$
represents the path integral
\begin{equation}
\int_{\tau_{\uparrow}}^{t}\mathrm{d}\boldsymbol{\boldsymbol{\tau}}=\sum_{m=2}^{\infty}\int_{\tau_{\uparrow}}^{t}\mathrm{d}\tau_{1}\int_{0}^{\tau_{1}}\mathrm{d}\tau_{2}...\int_{0}^{\tau_{m-1}}\mathrm{d}\tau_{m},
\end{equation}
which describes integration over the configuration subspace for which
at least one $\tau_{1}$ is within the interval $\left[\tau_{\uparrow},t\right]$.
This defines a formally exact inchworm step, which appears to leverage
knowledge of $\left\langle \sigma_{z}\left(\tau\right)\right\rangle $
for times up to $\tau_{\uparrow}$ in order to obtain the same observable
for the final time $t$. Examples of diagrams appearing in this expansion
are shown in Fig.~\ref{fig:cumulant_naive_inchworm}. Diagrams in
which the rightmost time index is to the left of $\tau_{\uparrow}$
(crossed out in the figure) are included in the $0^{\mathrm{th}}$
order contribution (diagram (1) in Fig.~\ref{fig:cumulant_naive_inchworm})
and need not be summed.

Unfortunately, the inchworm step we have just described cannot be
implemented as it stands, and has been introduced chiefly for didactic
purposes. This is because it includes contributions where $\left\langle \sigma_{z}\left(\tau\right)\right\rangle $
is needed at time argument $\tau>\tau_{\uparrow}$. Two examples are
overlaid with a question mark in Fig.~\ref{fig:cumulant_naive_inchworm}.
Such contributions are unknown and must be dropped from the expansion,
leading to an error the magnitude of which can be shown to be linear
in $\Delta t=t-\tau_{\uparrow}$. In practice, this makes convergence
of the algorithm to the exact result (by progressively reducing the
size of the inching time step $\Delta t$) very hard to achieve. However,
as Subsection~\ref{subsec:Cumulant-inchworm-algorithm} shows, this
issue can be overcome by taking a closer look at the structure of
the diagrams.

\subsection{Cumulant inchworm algorithm\label{subsec:Cumulant-inchworm-algorithm}}

\begin{figure}
\begin{raggedright}
(a) unwound
\par\end{raggedright}
\begin{centering}
\includegraphics{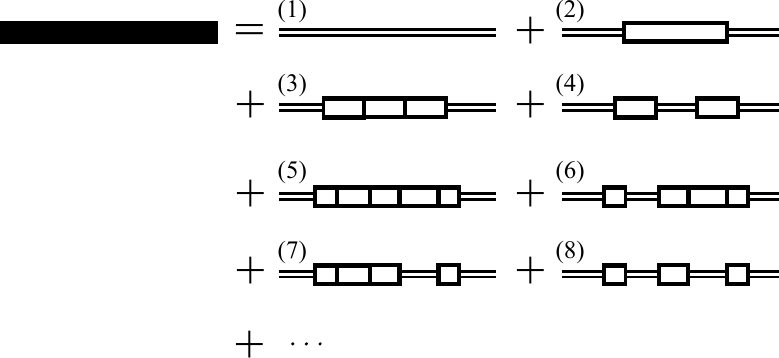}
\par\end{centering}
\begin{raggedright}
(b) inchworm
\par\end{raggedright}
\begin{centering}
\includegraphics{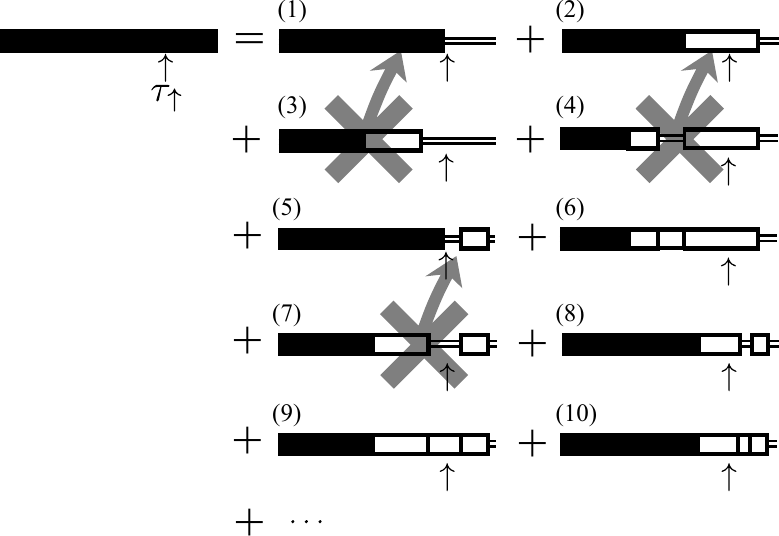}
\par\end{centering}
\caption{(a) The unwound dQMC expression for the full cumulant expansion. The
thick solid lines are the exact dynamics of expectation value and
the thin double lines are the unperturbed value $1$ within the diabatic
expansion. The solid-edged boxes with vertical ticks are the COP cumulants
as shown in Fig.~(\ref{fig:moment_and_cumulant}). Each configuration
may yield more than one diagram: $2$-time configurations gives one
$2^{\mathrm{nd}}$ order diagram (a.2); $4$-time configurations yield
diagrams (a.3) and (a.4) corresponding to 2 partitions in ${\cal P}_{4}$;
$6$-time configurations contain diagrams (a.5)\textendash (a.8) corresponding
to 4 partitions in ${\cal P}_{6}$. (b) The cumulant inchworm algorithm.
Any diagram that has a stand alone part (a cumulant box) to the left
of the $\uparrow$ has been included in the other diagrams and needs
to be neglected in the inchworm step. Diagram (b.3) is included in
diagram (b.1); diagram (b.4) is included in diagram (b.2); diagram
(b.7) is included in diagram (b.5). \label{fig:cumulant_improved_inchworm}}
\end{figure}

It is now necessary to solve the problem raised in Subsection~\ref{subsec:Naive-inchworm-algorithm},
namely the fact that one is unable to evaluate diagrams from configurations
having $\tau_{m}>\tau_{\uparrow}$ for some $m$. To do so, it is
possible to first \emph{unwind} the resumming done implicitly by the
cumulant expansion, then reintroduce it wherever possible. To see
how this works, one inserts the functional Eqs.~(\ref{eq:cumulant_expansion_integral})
and (\ref{eq:cumulant_expansion_functional_withP}). This gives
\begin{equation}
\begin{split}\left\langle \sigma_{z}\left(t\right)\right\rangle = & 1+\int_{0}^{t}\mathrm{d}\boldsymbol{\boldsymbol{\tau}}\gamma_{m}\left(\boldsymbol{\boldsymbol{\tau}}\right)\\
 & \ +\int_{0}^{t}\mathrm{d}\boldsymbol{\boldsymbol{\tau}}\gamma_{m}\left(\boldsymbol{\boldsymbol{\tau}}\right)\int_{0}^{\tau_{m}}\mathrm{d}\boldsymbol{\boldsymbol{\tau}}^{\prime}\mathcal{K}\left(\boldsymbol{\boldsymbol{\tau}}^{\prime}\right),
\end{split}
\end{equation}
and we sample an additional configuration $\boldsymbol{\boldsymbol{\tau}}^{\prime}$
for the integration $\int_{0}^{\tau_{m}}\mathrm{d}\boldsymbol{\boldsymbol{\tau}}^{\prime}$.
This can be iterated any number of times, generating an expansion
in terms of multiple cumulants, with the population pushed to increasingly
high-order terms. Examples of terms appearing in this unwound cumulant
expansion are shown in Fig.~\ref{fig:cumulant_improved_inchworm}a.
The term ``wound / wind'' is used to distinguish this procedure
from ``dressing / dress'' used in the context of Dyson equations,\cite{Fetter2003QuantumTheoryMany}
and in particular to distinguish ``unwound'' from ``bare.''

The unwound expansion can be written as
\begin{equation}
\left\langle \sigma_{z}\left(t\right)\right\rangle =1+\int\mathrm{d}\boldsymbol{\boldsymbol{\tau}}\Gamma\left(\boldsymbol{\boldsymbol{\tau}}\right),\label{eq:cumulant_expansion_integral_full}
\end{equation}
where the functional $\Gamma$ depends only on the COP cumulants.
At a general (even) order $m$, $\Gamma$ contains terms of various
partitions ${\cal P}_{m}$, as introduced in Sec.~\ref{subsec:Cumulants}:
\begin{equation}
\Gamma\left(\tau_{1},...,\tau_{m}\right)=\sum_{p\in{\cal P}_{m}}\prod_{\left(i_{1},i_{2},\cdots,i_{2n}\right)\in p}\gamma_{2n}\left(\tau_{i_{1}},\tau_{i_{2}}\cdots,\tau_{i_{2n}}\right).
\end{equation}
For instance, the lowest order functional ($m=2$) is simply
\begin{equation}
\Gamma\left(\tau_{1},\tau_{2}\right)=\gamma_{2}\left(\tau_{1},\tau_{2}\right),
\end{equation}
while that with $m=4$ contains two terms originating from the iteration
procedure: 
\begin{equation}
\Gamma\left(\tau_{1},\tau_{2},\tau_{3},\tau_{4}\right)=\gamma_{4}\left(\tau_{1},\tau_{2},\tau_{3},\tau_{4}\right)+\gamma_{2}\left(\tau_{1},\tau_{2}\right)\gamma_{2}\left(\tau_{3},\tau_{4}\right).
\end{equation}
Unlike the original bare expansion in diabatic coupling, each configuration
now generates multiple diagrams (corresponding to partitions). For
instance, as we show in Fig.~\ref{fig:cumulant_improved_inchworm}a,
a $4^{\mathrm{th}}$ order configuration generates $2$ diagrams,
(a.3) and (a.4), and a $6^{\mathrm{th}}$ order configuration generates
$4$ diagrams, (a.5)\textendash (a.8). We note briefly that it is
easy to show that the unwound expansion corresponds exactly to the
moment expansion, in the sense that $\Gamma_{i}=\mu_{i}$. However,
the advantages of the unwound representation will immediately become
apparent.

The unwinding completely removes the dependence on the population
$\left\langle \sigma_{z}\left(\tau\right)\right\rangle $, but does
so at the cost that the resummation properties of the COP expansion
are lost. We now \emph{partially rewind} the series wherever this
does not interfere with the assumptions of the inchworm step, in particular
the fact that we only have access to populations for $\tau<\tau_{\uparrow}$.
The inchworm step is performed by stochastically sampling configurations
$\boldsymbol{\tau}=\left(\tau_{1},...,\tau_{m}\right)\in\left[0,t\right]$,
as before
\begin{equation}
\left\langle \sigma_{z}\left(t\right)\right\rangle =\left\langle \sigma_{z}\left(\tau_{\uparrow}\right)\right\rangle +\int_{\tau_{\uparrow}}^{t}\mathrm{d}\boldsymbol{\boldsymbol{\tau}}\mathcal{K}^{\prime}\left(\boldsymbol{\boldsymbol{\tau}}\right).
\end{equation}
For each configuration, one sums only diagrams stemming from a \emph{proper}
subset of the partitions, ${\cal P}_{m}^{\prime}\subseteq{\cal P}_{m}$,
obtained by excluding partitions with subsequences (parts) having
all times in $\left[0,\tau_{\uparrow}\right]$. With this, we define
\begin{equation}
\Gamma^{\prime}\left(\boldsymbol{\tau}\right)=\sum_{p\in{\cal P}_{m}^{\prime}}\prod_{\left(i_{1},i_{2},\cdots,i_{2n}\right)\in p}\gamma_{2n}\left(\tau_{i_{1}},\tau_{i_{2}}\cdots,\tau_{i_{2n}}\right),\label{eq:cumulant_inchworm_functional-1}
\end{equation}
such that the functional to be summed is
\begin{equation}
{\cal K}^{\prime}\left(\boldsymbol{\tau}\right)=\begin{cases}
\Gamma^{\prime}\left(\boldsymbol{\tau}\right)\left\langle \sigma_{z}\left(\tau_{m}\right)\right\rangle  & \text{if }\tau_{m}<\tau_{\uparrow},\\
\Gamma^{\prime}\left(\boldsymbol{\tau}\right)\left\langle \sigma_{z}\left(\tau_{\uparrow}\right)\right\rangle  & \text{if }\tau_{m}>\tau_{\uparrow}.
\end{cases}\label{eq:cumulant_inchworm_functional-2}
\end{equation}
The diagrammatic representation of this cumulant inchworm expansion
is illustrated in Fig.~\ref{fig:cumulant_improved_inchworm}b, where
three examples of improper partitions (diagrams) are crossed out.
Note that contribution (b.8) takes into account precisely the kind
of diagram missing in the naive inchworm algorithm.

To justify that the cumulant inchworm expansion is formally equivalent
to the unwound expansion, it must be shown that the two sets of diagrams
generated by respective expansions are identical. To do so, we have
to prove that (a) these two sets of diagrams contain each other, in
the sense that every unwound diagram in one set is \emph{represented}
in the other; and (b), each diagram in one set is mapped to only a
\emph{single} diagram in the other set (such that the measure is conserved
under summation). We will proceed by example, rather than presenting
a formal proof.

To argue point (a), we need to show a containment relationship in
both directions. First, any cumulant inchworm diagram generates only
diagrams contained in the set of unwound diagrams. This is trivial
since the thick solid segment in each cumulant inchworm diagram can
be considered an infinite sum of unwound diagrams within that segment.
In the reverse direction, any unwound diagram can be found in the
set of cumulant inchworm diagrams: given an unwound diagram, one can
construct a cumulant inchworm diagram containing it by Eqs.~(\ref{eq:cumulant_inchworm_functional-1})
and (\ref{eq:cumulant_inchworm_functional-2}). As an example, we
consider the lowest order in Fig.~\ref{fig:cumulant_improved_inchworm}b
with a $2^{\mathrm{nd}}$ order configuration $\boldsymbol{\tau}=\left(\tau_{1},\tau_{2}\right)$.
The configuration generates one unwound diagram of the (a.2) type.
For the same configuration's cumulant inchworm diagram, three cases
are possible: $\tau_{1}>\tau_{\uparrow}>\tau_{2}$, $\tau_{\uparrow}>\tau_{1}>\tau_{2}$,
and $\tau_{1}>\tau_{2}>\tau_{\uparrow}$, which correspond to diagrams
(b.2), (b.3), and (b.5), respectively. It is clear that diagram (b.3)
is improper and has been included in diagram (b.1). Thus, an unwound
diagram of the (a.2) type is contained in (b.2), (b.3), or (b.1) depending
on its relationship with $\tau_{\uparrow}$.

Point (b) requires unique correspondence in both directions. One direction
is trivial\textemdash each cumulant inchworm diagram can be written
as an infinite sum of unique unwound diagrams. In the other direction,
we need to show that if there exist two cumulant inchworm diagrams
which contain the same unwound diagram, one of these two cumulant
inchworm diagrams must be eliminated. The propriety of cumulant inchworm
diagrams ensures this uniqueness: consider a $4^{\mathrm{th}}$-order
unwound diagram of type (a.4) with configuration $\boldsymbol{\tau}=\left(\tau_{1},\tau_{2},\tau_{3},\tau_{4}\right)$.
If $\tau_{1}>\tau_{\uparrow}>\tau_{2}$, the unwound diagram could
in principle be contained in two (not necessarily proper) cumulant
inchworm diagrams, (b.2) and (b.4). Diagram (b.4) is then eliminated
by the requirement of propriety. Similarly, if $\tau_{2}>\tau_{\uparrow}>\tau_{3}$,
the unwound diagram could be contained in two cumulant inchworm diagrams,
(b.5) and (b.7), but (b.7) is improper and therefore can be eliminated.
For other cases, the uniqueness is trivial: there is only one (necessarily
proper) cumulant inchworm diagram containing the unwound diagram.
For example, if $\tau_{3}>\tau_{\uparrow}>\tau_{4}$, only diagram
(b.8) can contain it.

With points (a) and (b) justified, it is clear that an exact correspondence
exists between the cumulant inchworm expansion and the unwound expansion.
Every diagram in the cumulant inchworm expansion corresponds to an
infinite number of unwound diagrams, and while the expansion does
not perform resummation over the entire length of the contour like
the system\textendash bath coupling expansion, it also has the distinct
advantage of scaling linearly in time. It therefore constitutes a
highly efficient method which is complementary to the system\textendash bath
coupling inchworm approach. We note in passing that a similar (cumulant-based)
approach to the system\textendash bath coupling expansion is possible,
but our preliminary attempts to pursue it indicated that it may in
practice be less efficient than the SBCI.

This method will be referred to as the Diabatic Coupling Cumulant
Inchworm (DCCI) approach in the companion paper.

\section{Conclusions\label{sec:Conclusions}}

In this paper we develop two complementary dQMC inchworm approaches
for the simulation of exact real-time non-adiabatic dynamics. These
approaches are based on generic expansions in either the system\textendash bath
coupling or the diabatic coupling, respectively, and thus should be
of general utility. For concreteness, as well as to permit benchmarking
of the approach, we specialize to the case of the spin\textendash boson
model. A discussion of the relative benefits and drawbacks of our
approach and detailed comparisons to established exact methods will
be presented in the companion paper. 

Our first approach is based on a system\textendash bath coupling
expansion, analogous to the hybridization expansion in the Anderson
model. Indeed the scheme is nearly identical to that employed in original
inchworm algorithm formulated for the Anderson impurity model.\cite{Cohen2015Tamingdynamicalsign}
Only the definitions of connectedness and propriety of diagrams differ,
due to the differences in the model systems studied. We formally show
that proper inchworm diagrams account for any diagram in the bare
Monte Carlo expansion once and only once. The major advantages of
the SBCI approach are twofold: there are far fewer proper inchworm
diagrams than bare diagrams and an infinite number of bare diagrams
are resummed in each contribution to the inchworm expansion. However,
this advantage comes at a cost, namely one has to calculate two-time
restricted propagators and perform a more involved error analysis
during inchworm propagation.

The second inchworm approach is based on the diabatic coupling expansion
and its cumulant form. Due to the fact that diagrams within the diabatic
coupling expansion include an interaction line between every two vertices,
the main advantages of the inchworm algorithm are lost if one follows
the previous scheme. To circumvent this problem, we introduce a cumulant
form of the expansion and propose an alternative inchworm approach,
the diabatic coupling cumulant inchworm (DCCI) expansion. The DCCI
expansion has the notable advantage that only single-time properties
are needed, and the simulation scales linearly in time. Since cumulant
forms can also be used in other inchworm expansions (such as the SBCI
approach), this property should be of general utility. We also note
that since the DCCI and SBCI expansions converge differently in distinct
parameter regimes, we expect their combined use to cover much, if
not all, of the relevant parameter space.

The formulations of Monte Carlo approaches with suppressed sign errors
presented here are quite general. They provide a framework allowing
for the simulation of both non-equilibrium and equilibrium observables
in general impurity-type problems. Many of these problems are out
of reach for existing methods and their exact interrogation remains
an open challenge. This work paves the way to explore the scope of
the inchworm dQMC methodology in such frontier problems, which will
be the subject of future studies.
\begin{acknowledgments}
We would like to thank Eran Rabani and Andrés Montoya-Castillo for
discussions. GC and DRR would like to thank Andrew J. Millis and Emanuel
Gull for previous collaboration on the inchworm algorithm. DRR acknowledges
support from NSF CHE-1464802. GC and HTC acknowledge support from
the Raymond and Beverly Sackler Center for Computational Molecular
and Materials Science, Tel Aviv University.

\end{acknowledgments}

\appendix

\section{Wick's theorem in the diabatic coupling expansion\label{sec:Wick's-Theorem-for}}

The multi-time correlation function of polaron shift operators given
in Eq.~(\ref{eq:multi-time-polaron-correlation}) can be written
as 
\begin{equation}
{\cal J}\left(\boldsymbol{s}\right)=\Bigl\langle\widetilde{{\cal B}}_{-}\left(s_{m+1}\right)\widetilde{{\cal B}}_{+}^{2}\left(s_{m}\right)\cdots\widetilde{{\cal B}}_{-}^{2}\left(s_{1}\right)\widetilde{{\cal B}}_{+}\left(s_{0}\right)\Bigr\rangle_{b}.
\end{equation}
The explicit form of the polaron shift operator in the interaction
picture is given by
\begin{align}
{\cal \widetilde{{\cal B}}}_{\sigma}\left(s\right)= & e^{\widetilde{\theta}_{\sigma}\left(s\right)},\\
\widetilde{\theta}_{\sigma}\left(s\right)= & \sigma\sum_{\ell}\frac{c_{\ell}}{\omega_{\ell}^{3/2}}\left(e^{i\omega_{\ell}s}b_{\ell}^{\dagger}-e^{-i\omega_{\ell}s}b_{\ell}\right).
\end{align}
To simplify the notation, we drop the $\ell$ index for the time being
and define $\xi\left(s\right)=\frac{c}{\omega^{3/2}}e^{i\omega s}$,
such that
\begin{equation}
\widetilde{\theta}_{\sigma}\left(s\right)=\sigma\left(\xi\left(s\right)b^{\dagger}-\xi\left(s\right)^{*}b\right).
\end{equation}
The arguments of the polaron shift operators product can be combined
using the identity
\begin{equation}
e^{vb^{\dagger}-v^{*}b}e^{ub^{\dagger}-u^{*}b}=e^{(v+u)b^{\dagger}-(v^{*}+u^{*})b}\times e^{(vu^{*}-v^{*}u)/2}
\end{equation}
for boson operators $b$ and $b^{\dagger}$ (as can easily be derived
using the Baker\textendash Campbell\textendash Hausdorff formula\cite{Messiah1961QuantumMechanics}).

Next, the two-time correlator of polaron shift operators is
\begin{equation}
\begin{split}\mathcal{B}_{\sigma^{\prime}}^{r^{\prime}}\left(s^{\prime}\right)\mathcal{B}_{\sigma}^{r}\left(s\right)= & \exp\left\{ \left[\sigma^{\prime}r^{\prime}\xi\left(s^{\prime}\right)+\sigma r\xi\left(s\right)\right]b^{\dagger}-\mathrm{c.c.}\right\} \times\\
 & \exp\left\{ i\sigma^{\prime}\sigma r^{\prime}r\mathrm{Im}\left[\xi\left(s^{\prime}\right)\xi\left(s\right)^{*}\right]\right\} .
\end{split}
\end{equation}
We note that the boson operator part of the correlator takes the same
form as the polaron shift operator and an additional scalar factor
(not a boson operator) emerges. Therefore, we can recursively combine
the argument using the above identity and find a general expression
for the multi-time correlator
\begin{equation}
\begin{split}\prod_{j}\mathcal{B}_{\sigma_{j}}^{r_{j}}\left(s_{j}\right)= & \exp\left\{ \sum_{j}\sigma_{j}r_{j}\xi\left(s_{j}\right)b^{\dagger}-\mathrm{c.c.}\right\} \times\\
 & \exp\left\{ i\sum_{j}\sum_{k<j}\sigma_{j}\sigma_{k}r_{j}r_{k}\mathrm{Im}\left[\xi\left(s_{j}\right)\xi\left(s_{k}\right)^{*}\right]\right\} .
\end{split}
\end{equation}
The scalar factor part can be rewritten in the form $\mathrm{Im}\left[\xi\left(s^{\prime}\right)\xi\left(s\right)^{*}\right]=\frac{c^{2}}{\omega^{3}}\sin\omega\left(s^{\prime}-s\right)$. 

We now focus on the thermal average of the boson operator, $\exp\left\{ \sum_{j}\sigma_{j}r_{j}\xi\left(s_{j}\right)b^{\dagger}-\mathrm{c.c.}\right\} $.
The thermal average of free boson operator of the above form can be
obtained as
\begin{equation}
\mathrm{Tr}_{b}\left\{ \rho_{b}e^{\kappa b^{\dagger}-\kappa^{*}b}\right\} =\exp\left\{ -\frac{1}{2}\kappa\kappa^{*}\coth\left(\frac{\beta\omega}{2}\right)\right\} ,
\end{equation}
where $\rho_{b}=e^{-\beta H_{b}}$. We can take the thermal average
of the two-time correlator to obtain 
\begin{eqnarray}
 &  & \mathrm{Tr}_{b}\left\{ \rho_{b}\mathcal{B}_{\sigma^{\prime}}^{r^{\prime}}\left(s^{\prime}\right)\mathcal{B}_{\sigma}^{r}\left(s\right)\right\} =\nonumber \\
 &  & \ \exp\left\{ \sigma^{\prime}\sigma r^{\prime}r\frac{c^{2}}{\omega^{3}}\left[1-\cos\omega\left(s^{\prime}-s\right)\right]\coth\left(\frac{\beta\omega}{2}\right)\right\} \times\nonumber \\
 &  & \ \exp\left\{ i\sigma^{\prime}\sigma r^{\prime}r\frac{c^{2}}{\omega^{3}}\sin\omega\left(s^{\prime}-s\right)\right\} ,
\end{eqnarray}
where a time-independent phase is dropped, since it cancels out when
a configuration on the Keldysh contour is considered. By choosing
$\sigma^{\prime}=-1$ and $\sigma=1$ and putting the $\ell$ index
back, we can carry out $\sum_{\ell}$ in terms of spectral density
$J\left(\omega\right)$,
\begin{equation}
\begin{split} & \mathrm{Tr}_{b}\left\{ \rho_{b}\mathcal{B}_{-}^{r^{\prime}}\left(s^{\prime}\right)\mathcal{B}_{+}^{r}\left(s\right)\right\} =\\
 & \qquad\exp\left\{ r^{\prime}r\left[-\mathcal{Q}_{2}\left(s^{\prime}-s\right)-i\mathcal{Q}_{1}\left(s^{\prime}-s\right)\right]\right\} ,
\end{split}
\end{equation}
where $\mathcal{Q}_{1}$and $\mathcal{Q}_{2}$ are defined by Eqs.~(\ref{eq:Q1})
and (\ref{eq:Q2}). The two-time correlation function, Eq.~(\ref{eq:correlation_function_inQ1Q2}),
is then given by
\begin{equation}
C\left(s^{\prime},s\right)^{r^{\prime}r}\equiv\mathrm{Tr}_{b}\left\{ \rho_{b}\mathcal{B}_{-}^{r^{\prime}}\left(s^{\prime}\right)\mathcal{B}_{+}^{r}\left(s\right)\right\} 
\end{equation}

Finally, we take the thermal average of the multi-time correlator

\begin{widetext}
\begin{equation}
\begin{split}
\mathrm{Tr}_{b}\left\{ \rho_{b}\prod_{j}\mathcal{B}_{\sigma_{j}}^{r_{j}}\left(s_{j}\right)\right\} = & \exp\left\{ \sum_{j}\sum_{k<j}\sigma_{j}\sigma_{k}r_{j}r_{k}\frac{c^{2}}{\omega^{3}}\left[1-\cos\omega\left(s_{j}-s_{k}\right)\right]\coth\left(\frac{\beta\omega}{2}\right)\right\} \times\\ & \exp\left\{ i\sum_{j}\sum_{k<j}\sigma_{j}\sigma_{k}r_{j}r_{k}\frac{c^{2}}{\omega^{3}}\sin\omega\left(s_{j}-s_{k}\right)\right\} .\end{split}
\end{equation}
\end{widetext}We can finally carry out $\sum_{\ell}$ in terms of spectral density
$J\left(\omega\right)$, concluding that
\begin{equation}
\mathrm{Tr}_{b}\left\{ \rho_{b}\prod_{j}\mathcal{B}_{\sigma_{j}}^{r_{j}}\left(s_{j}\right)\right\} =\prod_{j}\prod_{k<j}\left(C\left(s_{j},s_{k}\right)^{r_{j}r_{k}}\right)^{-\sigma_{j}\sigma_{k}}.
\end{equation}
Since we have $\sigma_{j}=1$ for $j$ even and $\sigma_{j}=-1$ for
$j$ odd, the powers are
\begin{equation}
-\sigma_{j}\sigma_{k}=\begin{cases}
1 & \left|j-k\right|\ \mathrm{odd}\\
-1 & \left|j-k\right|\ \mathrm{even}
\end{cases}.
\end{equation}

\bibliographystyle{aipnum4-1}
\bibliography{../bibtex/dqmc_spin_boson}

\begin{thebibliography}{63}%
\makeatletter
\providecommand \@ifxundefined [1]{%
 \@ifx{#1\undefined}
}%
\providecommand \@ifnum [1]{%
 \ifnum #1\expandafter \@firstoftwo
 \else \expandafter \@secondoftwo
 \fi
}%
\providecommand \@ifx [1]{%
 \ifx #1\expandafter \@firstoftwo
 \else \expandafter \@secondoftwo
 \fi
}%
\providecommand \natexlab [1]{#1}%
\providecommand \enquote  [1]{``#1''}%
\providecommand \bibnamefont  [1]{#1}%
\providecommand \bibfnamefont [1]{#1}%
\providecommand \citenamefont [1]{#1}%
\providecommand \href@noop [0]{\@secondoftwo}%
\providecommand \href [0]{\begingroup \@sanitize@url \@href}%
\providecommand \@href[1]{\@@startlink{#1}\@@href}%
\providecommand \@@href[1]{\endgroup#1\@@endlink}%
\providecommand \@sanitize@url [0]{\catcode `\\12\catcode `\$12\catcode
  `\&12\catcode `\#12\catcode `\^12\catcode `\_12\catcode `\%12\relax}%
\providecommand \@@startlink[1]{}%
\providecommand \@@endlink[0]{}%
\providecommand \url  [0]{\begingroup\@sanitize@url \@url }%
\providecommand \@url [1]{\endgroup\@href {#1}{\urlprefix }}%
\providecommand \urlprefix  [0]{URL }%
\providecommand \Eprint [0]{\href }%
\providecommand \doibase [0]{http://dx.doi.org/}%
\providecommand \selectlanguage [0]{\@gobble}%
\providecommand \bibinfo  [0]{\@secondoftwo}%
\providecommand \bibfield  [0]{\@secondoftwo}%
\providecommand \translation [1]{[#1]}%
\providecommand \BibitemOpen [0]{}%
\providecommand \bibitemStop [0]{}%
\providecommand \bibitemNoStop [0]{.\EOS\space}%
\providecommand \EOS [0]{\spacefactor3000\relax}%
\providecommand \BibitemShut  [1]{\csname bibitem#1\endcsname}%
\let\auto@bib@innerbib\@empty
\bibitem [{\citenamefont {Nitzan}\ and\ \citenamefont
  {Ratner}(2003)}]{Nitzan2003Electrontransportin}%
  \BibitemOpen
  \bibfield  {author} {\bibinfo {author} {\bibfnamefont {A.}~\bibnamefont
  {Nitzan}}\ and\ \bibinfo {author} {\bibfnamefont {M.~A.}\ \bibnamefont
  {Ratner}},\ }\href {\doibase 10.1126/science.1081572} {\bibfield  {journal}
  {\bibinfo  {journal} {Science}\ }\textbf {\bibinfo {volume} {300}},\ \bibinfo
  {pages} {1384} (\bibinfo {year} {2003})}\BibitemShut {NoStop}%
\bibitem [{\citenamefont {Fausti}\ \emph {et~al.}(2011)\citenamefont {Fausti},
  \citenamefont {Tobey}, \citenamefont {Dean}, \citenamefont {Kaiser},
  \citenamefont {Dienst}, \citenamefont {Hoffmann}, \citenamefont {Pyon},
  \citenamefont {Takayama}, \citenamefont {Takagi},\ and\ \citenamefont
  {Cavalleri}}]{Fausti2011Lightinducedsuperconductivity}%
  \BibitemOpen
  \bibfield  {author} {\bibinfo {author} {\bibfnamefont {D.}~\bibnamefont
  {Fausti}}, \bibinfo {author} {\bibfnamefont {R.}~\bibnamefont {Tobey}},
  \bibinfo {author} {\bibfnamefont {N.}~\bibnamefont {Dean}}, \bibinfo {author}
  {\bibfnamefont {S.}~\bibnamefont {Kaiser}}, \bibinfo {author} {\bibfnamefont
  {A.}~\bibnamefont {Dienst}}, \bibinfo {author} {\bibfnamefont {M.~C.}\
  \bibnamefont {Hoffmann}}, \bibinfo {author} {\bibfnamefont {S.}~\bibnamefont
  {Pyon}}, \bibinfo {author} {\bibfnamefont {T.}~\bibnamefont {Takayama}},
  \bibinfo {author} {\bibfnamefont {H.}~\bibnamefont {Takagi}}, \ and\ \bibinfo
  {author} {\bibfnamefont {A.}~\bibnamefont {Cavalleri}},\ }\href@noop {}
  {\bibfield  {journal} {\bibinfo  {journal} {Science}\ }\textbf {\bibinfo
  {volume} {331}},\ \bibinfo {pages} {189} (\bibinfo {year}
  {2011})}\BibitemShut {NoStop}%
\bibitem [{\citenamefont {Negele}\ and\ \citenamefont
  {Orland}(1988)}]{Negele1988Quantummanyparticle}%
  \BibitemOpen
  \bibfield  {author} {\bibinfo {author} {\bibfnamefont {J.~W.}\ \bibnamefont
  {Negele}}\ and\ \bibinfo {author} {\bibfnamefont {H.}~\bibnamefont
  {Orland}},\ }\href@noop {} {\emph {\bibinfo {title} {Quantum many-particle
  systems}}},\ Vol.\ \bibinfo {volume} {200}\ (\bibinfo  {publisher}
  {Addison-Wesley New York},\ \bibinfo {year} {1988})\BibitemShut {NoStop}%
\bibitem [{\citenamefont {Bloch}, \citenamefont {Dalibard},\ and\ \citenamefont
  {Zwerger}(2008)}]{Bloch2008Manybodyphysics}%
  \BibitemOpen
  \bibfield  {author} {\bibinfo {author} {\bibfnamefont {I.}~\bibnamefont
  {Bloch}}, \bibinfo {author} {\bibfnamefont {J.}~\bibnamefont {Dalibard}}, \
  and\ \bibinfo {author} {\bibfnamefont {W.}~\bibnamefont {Zwerger}},\ }\href
  {\doibase 10.1103/RevModPhys.80.885} {\bibfield  {journal} {\bibinfo
  {journal} {Rev. Mod. Phys.}\ }\textbf {\bibinfo {volume} {80}},\ \bibinfo
  {pages} {885} (\bibinfo {year} {2008})}\BibitemShut {NoStop}%
\bibitem [{\citenamefont {Trotzky}\ \emph {et~al.}(2010)\citenamefont
  {Trotzky}, \citenamefont {Pollet}, \citenamefont {Gerbier}, \citenamefont
  {Schnorrberger}, \citenamefont {Bloch}, \citenamefont {Prokof'ev},
  \citenamefont {Svistunov},\ and\ \citenamefont
  {Troyer}}]{Trotzky2010Suppressioncriticaltemperature}%
  \BibitemOpen
  \bibfield  {author} {\bibinfo {author} {\bibfnamefont {S.}~\bibnamefont
  {Trotzky}}, \bibinfo {author} {\bibfnamefont {L.}~\bibnamefont {Pollet}},
  \bibinfo {author} {\bibfnamefont {F.}~\bibnamefont {Gerbier}}, \bibinfo
  {author} {\bibfnamefont {U.}~\bibnamefont {Schnorrberger}}, \bibinfo {author}
  {\bibfnamefont {I.}~\bibnamefont {Bloch}}, \bibinfo {author} {\bibfnamefont
  {N.~V.}\ \bibnamefont {Prokof'ev}}, \bibinfo {author} {\bibfnamefont
  {B.}~\bibnamefont {Svistunov}}, \ and\ \bibinfo {author} {\bibfnamefont
  {M.}~\bibnamefont {Troyer}},\ }\href {\doibase 10.1038/nphys1799} {\bibfield
  {journal} {\bibinfo  {journal} {Nature Physics}\ }\textbf {\bibinfo {volume}
  {6}},\ \bibinfo {pages} {998} (\bibinfo {year} {2010})}\BibitemShut {NoStop}%
\bibitem [{\citenamefont {Kashurnikov}, \citenamefont {Prokof'ev},\ and\
  \citenamefont
  {Svistunov}(2002)}]{Kashurnikov2002Revealingsuperfluidchar21Mottinsulator}%
  \BibitemOpen
  \bibfield  {author} {\bibinfo {author} {\bibfnamefont {V.~A.}\ \bibnamefont
  {Kashurnikov}}, \bibinfo {author} {\bibfnamefont {N.~V.}\ \bibnamefont
  {Prokof'ev}}, \ and\ \bibinfo {author} {\bibfnamefont {B.~V.}\ \bibnamefont
  {Svistunov}},\ }\href {\doibase 10.1103/PhysRevA.66.031601} {\bibfield
  {journal} {\bibinfo  {journal} {Phys. Rev. A}\ }\textbf {\bibinfo {volume}
  {66}},\ \bibinfo {pages} {031601} (\bibinfo {year} {2002})}\BibitemShut
  {NoStop}%
\bibitem [{\citenamefont {Troyer}\ and\ \citenamefont
  {Wiese}(2005)}]{Troyer2005ComputationalComplexityand}%
  \BibitemOpen
  \bibfield  {author} {\bibinfo {author} {\bibfnamefont {M.}~\bibnamefont
  {Troyer}}\ and\ \bibinfo {author} {\bibfnamefont {U.-J.}\ \bibnamefont
  {Wiese}},\ }\href {\doibase 10.1103/PhysRevLett.94.170201} {\bibfield
  {journal} {\bibinfo  {journal} {Phys. Rev. Lett.}\ }\textbf {\bibinfo
  {volume} {94}},\ \bibinfo {pages} {170201} (\bibinfo {year}
  {2005})}\BibitemShut {NoStop}%
\bibitem [{\citenamefont {Shakhnovich}\ \emph {et~al.}(1991)\citenamefont
  {Shakhnovich}, \citenamefont {Farztdinov}, \citenamefont {Gutin},\ and\
  \citenamefont {Karplus}}]{Shakhnovich1991Proteinfoldingbottlenecks}%
  \BibitemOpen
  \bibfield  {author} {\bibinfo {author} {\bibfnamefont {E.}~\bibnamefont
  {Shakhnovich}}, \bibinfo {author} {\bibfnamefont {G.}~\bibnamefont
  {Farztdinov}}, \bibinfo {author} {\bibfnamefont {A.~M.}\ \bibnamefont
  {Gutin}}, \ and\ \bibinfo {author} {\bibfnamefont {M.}~\bibnamefont
  {Karplus}},\ }\href {\doibase 10.1103/PhysRevLett.67.1665} {\bibfield
  {journal} {\bibinfo  {journal} {Phys. Rev. Lett.}\ }\textbf {\bibinfo
  {volume} {67}},\ \bibinfo {pages} {1665} (\bibinfo {year}
  {1991})}\BibitemShut {NoStop}%
\bibitem [{\citenamefont {Saxton}(1996)}]{Saxton1996Anomalousdiffusiondue}%
  \BibitemOpen
  \bibfield  {author} {\bibinfo {author} {\bibfnamefont {M.~J.}\ \bibnamefont
  {Saxton}},\ }\href {http://www.ncbi.nlm.nih.gov/pmc/articles/PMC1225051/}
  {\bibfield  {journal} {\bibinfo  {journal} {Biophys. J.}\ }\textbf {\bibinfo
  {volume} {70}},\ \bibinfo {pages} {1250} (\bibinfo {year}
  {1996})}\BibitemShut {NoStop}%
\bibitem [{\citenamefont {Prokof'ev}\ and\ \citenamefont
  {Svistunov}(2001)}]{Prokofev2001WormAlgorithmsClassical}%
  \BibitemOpen
  \bibfield  {author} {\bibinfo {author} {\bibfnamefont {N.}~\bibnamefont
  {Prokof'ev}}\ and\ \bibinfo {author} {\bibfnamefont {B.}~\bibnamefont
  {Svistunov}},\ }\href {\doibase 10.1103/PhysRevLett.87.160601} {\bibfield
  {journal} {\bibinfo  {journal} {Phys. Rev. Lett.}\ }\textbf {\bibinfo
  {volume} {87}},\ \bibinfo {pages} {160601} (\bibinfo {year}
  {2001})}\BibitemShut {NoStop}%
\bibitem [{\citenamefont {Newman}\ and\ \citenamefont
  {Barkema}(1999)}]{Newman1999MonteCarloMethods}%
  \BibitemOpen
  \bibfield  {author} {\bibinfo {author} {\bibfnamefont {M.~E.~J.}\
  \bibnamefont {Newman}}\ and\ \bibinfo {author} {\bibfnamefont {G.~T.}\
  \bibnamefont {Barkema}},\ }\href
  {http://itf.fys.kuleuven.be/fpspXIII/material/Barkema_FPSPXIII.pdf} {\emph
  {\bibinfo {title} {Monte {Carlo} {Methods} in {Statistical} {Physics}}}}\
  (\bibinfo  {publisher} {Oxford University Press: New York, USA},\ \bibinfo
  {year} {1999})\BibitemShut {NoStop}%
\bibitem [{\citenamefont {Mak}\ and\ \citenamefont
  {Chandler}(1990)}]{Mak1990Solvingsignproblem}%
  \BibitemOpen
  \bibfield  {author} {\bibinfo {author} {\bibfnamefont {C.~H.}\ \bibnamefont
  {Mak}}\ and\ \bibinfo {author} {\bibfnamefont {D.}~\bibnamefont {Chandler}},\
  }\href {\doibase 10.1103/PhysRevA.41.5709} {\bibfield  {journal} {\bibinfo
  {journal} {Phys. Rev. A}\ }\textbf {\bibinfo {volume} {41}},\ \bibinfo
  {pages} {5709} (\bibinfo {year} {1990})}\BibitemShut {NoStop}%
\bibitem [{\citenamefont {Burovski}\ \emph {et~al.}(2006)\citenamefont
  {Burovski}, \citenamefont {Prokof'ev}, \citenamefont {Svistunov},\ and\
  \citenamefont {Troyer}}]{Burovski2006CriticalTemperatureand}%
  \BibitemOpen
  \bibfield  {author} {\bibinfo {author} {\bibfnamefont {E.}~\bibnamefont
  {Burovski}}, \bibinfo {author} {\bibfnamefont {N.}~\bibnamefont {Prokof'ev}},
  \bibinfo {author} {\bibfnamefont {B.}~\bibnamefont {Svistunov}}, \ and\
  \bibinfo {author} {\bibfnamefont {M.}~\bibnamefont {Troyer}},\ }\href
  {\doibase 10.1103/PhysRevLett.96.160402} {\bibfield  {journal} {\bibinfo
  {journal} {Phys. Rev. Lett.}\ }\textbf {\bibinfo {volume} {96}},\ \bibinfo
  {pages} {160402} (\bibinfo {year} {2006})}\BibitemShut {NoStop}%
\bibitem [{\citenamefont {Hirsch}\ \emph {et~al.}(1982)\citenamefont {Hirsch},
  \citenamefont {Sugar}, \citenamefont {Scalapino},\ and\ \citenamefont
  {Blankenbecler}}]{Hirsch1982MonteCarlosimulations}%
  \BibitemOpen
  \bibfield  {author} {\bibinfo {author} {\bibfnamefont {J.~E.}\ \bibnamefont
  {Hirsch}}, \bibinfo {author} {\bibfnamefont {R.~L.}\ \bibnamefont {Sugar}},
  \bibinfo {author} {\bibfnamefont {D.~J.}\ \bibnamefont {Scalapino}}, \ and\
  \bibinfo {author} {\bibfnamefont {R.}~\bibnamefont {Blankenbecler}},\ }\href
  {\doibase 10.1103/PhysRevB.26.5033} {\bibfield  {journal} {\bibinfo
  {journal} {Phys. Rev. B}\ }\textbf {\bibinfo {volume} {26}},\ \bibinfo
  {pages} {5033} (\bibinfo {year} {1982})}\BibitemShut {NoStop}%
\bibitem [{\citenamefont {Hirsch}(1988)}]{Hirsch1988Stablemontecarlo}%
  \BibitemOpen
  \bibfield  {author} {\bibinfo {author} {\bibfnamefont {J.~E.}\ \bibnamefont
  {Hirsch}},\ }\href {\doibase 10.1103/PhysRevB.38.12023} {\bibfield  {journal}
  {\bibinfo  {journal} {Phys. Rev. B}\ }\textbf {\bibinfo {volume} {38}},\
  \bibinfo {pages} {12023} (\bibinfo {year} {1988})}\BibitemShut {NoStop}%
\bibitem [{\citenamefont {Egger}, \citenamefont {M{\"{u}}hlbacher},\ and\
  \citenamefont {Mak}(2000)}]{Egger2000PathintegralMonte}%
  \BibitemOpen
  \bibfield  {author} {\bibinfo {author} {\bibfnamefont {R.}~\bibnamefont
  {Egger}}, \bibinfo {author} {\bibfnamefont {L.}~\bibnamefont
  {M{\"{u}}hlbacher}}, \ and\ \bibinfo {author} {\bibfnamefont {C.~H.}\
  \bibnamefont {Mak}},\ }\href {\doibase 10.1103/PhysRevE.61.5961} {\bibfield
  {journal} {\bibinfo  {journal} {Phys. Rev. E}\ }\textbf {\bibinfo {volume}
  {61}},\ \bibinfo {pages} {5961} (\bibinfo {year} {2000})}\BibitemShut
  {NoStop}%
\bibitem [{\citenamefont {Maier}\ \emph {et~al.}(2005)\citenamefont {Maier},
  \citenamefont {Jarrell}, \citenamefont {Pruschke},\ and\ \citenamefont
  {Hettler}}]{Maier2005Quantumclustertheories}%
  \BibitemOpen
  \bibfield  {author} {\bibinfo {author} {\bibfnamefont {T.}~\bibnamefont
  {Maier}}, \bibinfo {author} {\bibfnamefont {M.}~\bibnamefont {Jarrell}},
  \bibinfo {author} {\bibfnamefont {T.}~\bibnamefont {Pruschke}}, \ and\
  \bibinfo {author} {\bibfnamefont {M.~H.}\ \bibnamefont {Hettler}},\ }\href
  {\doibase 10.1103/RevModPhys.77.1027} {\bibfield  {journal} {\bibinfo
  {journal} {Reviews of Modern Physics}\ }\textbf {\bibinfo {volume} {77}},\
  \bibinfo {pages} {1027} (\bibinfo {year} {2005})}\BibitemShut {NoStop}%
\bibitem [{\citenamefont {Needs}\ \emph {et~al.}(2010)\citenamefont {Needs},
  \citenamefont {Towler}, \citenamefont {Drummond},\ and\ \citenamefont
  {Ríos}}]{Needs2010Continuumvariationaland}%
  \BibitemOpen
  \bibfield  {author} {\bibinfo {author} {\bibfnamefont {R.~J.}\ \bibnamefont
  {Needs}}, \bibinfo {author} {\bibfnamefont {M.~D.}\ \bibnamefont {Towler}},
  \bibinfo {author} {\bibfnamefont {N.~D.}\ \bibnamefont {Drummond}}, \ and\
  \bibinfo {author} {\bibfnamefont {P.~L.}\ \bibnamefont {Ríos}},\ }\href
  {\doibase 10.1088/0953-8984/22/2/023201} {\bibfield  {journal} {\bibinfo
  {journal} {Journal of Physics: Condensed Matter}\ }\textbf {\bibinfo {volume}
  {22}},\ \bibinfo {pages} {023201} (\bibinfo {year} {2010})}\BibitemShut
  {NoStop}%
\bibitem [{\citenamefont {LeBlanc}\ \emph {et~al.}(2015)\citenamefont
  {LeBlanc}, \citenamefont {Antipov}, \citenamefont {Becca}, \citenamefont
  {Bulik}, \citenamefont {Chan}, \citenamefont {Chung}, \citenamefont {Deng},
  \citenamefont {Ferrero}, \citenamefont {Henderson}, \citenamefont
  {Jim\'enez-Hoyos}, \citenamefont {Kozik}, \citenamefont {Liu}, \citenamefont
  {Millis}, \citenamefont {Prokof'ev}, \citenamefont {Qin}, \citenamefont
  {Scuseria}, \citenamefont {Shi}, \citenamefont {Svistunov}, \citenamefont
  {Tocchio}, \citenamefont {Tupitsyn}, \citenamefont {White}, \citenamefont
  {Zhang}, \citenamefont {Zheng}, \citenamefont {Zhu},\ and\ \citenamefont
  {Gull}}]{LeBlanc2015SolutionsTwoDimensional}%
  \BibitemOpen
  \bibfield  {author} {\bibinfo {author} {\bibfnamefont {J.~P.~F.}\
  \bibnamefont {LeBlanc}}, \bibinfo {author} {\bibfnamefont {A.~E.}\
  \bibnamefont {Antipov}}, \bibinfo {author} {\bibfnamefont {F.}~\bibnamefont
  {Becca}}, \bibinfo {author} {\bibfnamefont {I.~W.}\ \bibnamefont {Bulik}},
  \bibinfo {author} {\bibfnamefont {G.~K.-L.}\ \bibnamefont {Chan}}, \bibinfo
  {author} {\bibfnamefont {C.-M.}\ \bibnamefont {Chung}}, \bibinfo {author}
  {\bibfnamefont {Y.}~\bibnamefont {Deng}}, \bibinfo {author} {\bibfnamefont
  {M.}~\bibnamefont {Ferrero}}, \bibinfo {author} {\bibfnamefont {T.~M.}\
  \bibnamefont {Henderson}}, \bibinfo {author} {\bibfnamefont {C.~A.}\
  \bibnamefont {Jim\'enez-Hoyos}}, \bibinfo {author} {\bibfnamefont
  {E.}~\bibnamefont {Kozik}}, \bibinfo {author} {\bibfnamefont {X.-W.}\
  \bibnamefont {Liu}}, \bibinfo {author} {\bibfnamefont {A.~J.}\ \bibnamefont
  {Millis}}, \bibinfo {author} {\bibfnamefont {N.~V.}\ \bibnamefont
  {Prokof'ev}}, \bibinfo {author} {\bibfnamefont {M.}~\bibnamefont {Qin}},
  \bibinfo {author} {\bibfnamefont {G.~E.}\ \bibnamefont {Scuseria}}, \bibinfo
  {author} {\bibfnamefont {H.}~\bibnamefont {Shi}}, \bibinfo {author}
  {\bibfnamefont {B.~V.}\ \bibnamefont {Svistunov}}, \bibinfo {author}
  {\bibfnamefont {L.~F.}\ \bibnamefont {Tocchio}}, \bibinfo {author}
  {\bibfnamefont {I.~S.}\ \bibnamefont {Tupitsyn}}, \bibinfo {author}
  {\bibfnamefont {S.~R.}\ \bibnamefont {White}}, \bibinfo {author}
  {\bibfnamefont {S.}~\bibnamefont {Zhang}}, \bibinfo {author} {\bibfnamefont
  {B.-X.}\ \bibnamefont {Zheng}}, \bibinfo {author} {\bibfnamefont
  {Z.}~\bibnamefont {Zhu}}, \ and\ \bibinfo {author} {\bibfnamefont
  {E.}~\bibnamefont {Gull}} (\bibinfo {collaboration} {Simons Collaboration on
  the Many-Electron Problem}),\ }\href {\doibase 10.1103/PhysRevX.5.041041}
  {\bibfield  {journal} {\bibinfo  {journal} {Phys. Rev. X}\ }\textbf {\bibinfo
  {volume} {5}},\ \bibinfo {pages} {041041} (\bibinfo {year}
  {2015})}\BibitemShut {NoStop}%
\bibitem [{\citenamefont {Rombouts}, \citenamefont {Heyde},\ and\ \citenamefont
  {Jachowicz}(1998)}]{Rombouts1998discreteHubbardStratonovich}%
  \BibitemOpen
  \bibfield  {author} {\bibinfo {author} {\bibfnamefont {S.}~\bibnamefont
  {Rombouts}}, \bibinfo {author} {\bibfnamefont {K.}~\bibnamefont {Heyde}}, \
  and\ \bibinfo {author} {\bibfnamefont {N.}~\bibnamefont {Jachowicz}},\ }\href
  {\doibase 10.1016/S0375-9601(98)00197-2} {\bibfield  {journal} {\bibinfo
  {journal} {Phys. Lett. A}\ }\textbf {\bibinfo {volume} {242}},\ \bibinfo
  {pages} {271} (\bibinfo {year} {1998})}\BibitemShut {NoStop}%
\bibitem [{\citenamefont {M{\"{u}}hlbacher}\ and\ \citenamefont
  {Rabani}(2008)}]{Muehlbacher2008Realtimepath}%
  \BibitemOpen
  \bibfield  {author} {\bibinfo {author} {\bibfnamefont {L.}~\bibnamefont
  {M{\"{u}}hlbacher}}\ and\ \bibinfo {author} {\bibfnamefont {E.}~\bibnamefont
  {Rabani}},\ }\href {http://link.aps.org/doi/10.1103/PhysRevLett.100.176403}
  {\bibfield  {journal} {\bibinfo  {journal} {Phys. Rev. Lett.}\ }\textbf
  {\bibinfo {volume} {100}},\ \bibinfo {pages} {176403} (\bibinfo {year}
  {2008})},\ \Eprint {http://arxiv.org/abs/0707.0956} {0707.0956} \BibitemShut
  {NoStop}%
\bibitem [{\citenamefont {Schir{\'{o}}}(2010)}]{Schiro2010Realtimedynamics}%
  \BibitemOpen
  \bibfield  {author} {\bibinfo {author} {\bibfnamefont {M.}~\bibnamefont
  {Schir{\'{o}}}},\ }\href {\doibase 10.1103/PhysRevB.81.085126} {\bibfield
  {journal} {\bibinfo  {journal} {Phys. Rev. B}\ }\textbf {\bibinfo {volume}
  {81}},\ \bibinfo {pages} {085126} (\bibinfo {year} {2010})}\BibitemShut
  {NoStop}%
\bibitem [{\citenamefont {Werner}\ \emph {et~al.}(2006)\citenamefont {Werner},
  \citenamefont {Comanac}, \citenamefont {{De' Medici}}, \citenamefont
  {Troyer}, \citenamefont {Millis}, \citenamefont {{De Medici}}, \citenamefont
  {Troyer},\ and\ \citenamefont {Millis}}]{Werner2006Continuoustimesolver}%
  \BibitemOpen
  \bibfield  {author} {\bibinfo {author} {\bibfnamefont {P.}~\bibnamefont
  {Werner}}, \bibinfo {author} {\bibfnamefont {A.}~\bibnamefont {Comanac}},
  \bibinfo {author} {\bibfnamefont {L.}~\bibnamefont {{De' Medici}}}, \bibinfo
  {author} {\bibfnamefont {M.}~\bibnamefont {Troyer}}, \bibinfo {author}
  {\bibfnamefont {A.~J.}\ \bibnamefont {Millis}}, \bibinfo {author}
  {\bibfnamefont {L.}~\bibnamefont {{De Medici}}}, \bibinfo {author}
  {\bibfnamefont {M.}~\bibnamefont {Troyer}}, \ and\ \bibinfo {author}
  {\bibfnamefont {A.~J.}\ \bibnamefont {Millis}},\ }\href
  {http://journals.aps.org/prl/abstract/10.1103/PhysRevLett.97.076405}
  {\bibfield  {journal} {\bibinfo  {journal} {Phys. Rev. Lett.}\ }\textbf
  {\bibinfo {volume} {97}},\ \bibinfo {pages} {076405} (\bibinfo {year}
  {2006})}\BibitemShut {NoStop}%
\bibitem [{\citenamefont {Werner}\ and\ \citenamefont
  {Millis}(2006)}]{Werner2006Hybridizationexpansionimpurity}%
  \BibitemOpen
  \bibfield  {author} {\bibinfo {author} {\bibfnamefont {P.}~\bibnamefont
  {Werner}}\ and\ \bibinfo {author} {\bibfnamefont {A.~J.}\ \bibnamefont
  {Millis}},\ }\href {\doibase 10.1103/PhysRevB.74.155107} {\bibfield
  {journal} {\bibinfo  {journal} {Phys. Rev. B}\ }\textbf {\bibinfo {volume}
  {74}},\ \bibinfo {pages} {155107} (\bibinfo {year} {2006})}\BibitemShut
  {NoStop}%
\bibitem [{\citenamefont {Gull}\ \emph {et~al.}(2008)\citenamefont {Gull},
  \citenamefont {Werner}, \citenamefont {Parcollet},\ and\ \citenamefont
  {Troyer}}]{Gull2008Continuoustimeauxiliary}%
  \BibitemOpen
  \bibfield  {author} {\bibinfo {author} {\bibfnamefont {E.}~\bibnamefont
  {Gull}}, \bibinfo {author} {\bibfnamefont {P.}~\bibnamefont {Werner}},
  \bibinfo {author} {\bibfnamefont {O.}~\bibnamefont {Parcollet}}, \ and\
  \bibinfo {author} {\bibfnamefont {M.}~\bibnamefont {Troyer}},\ }\href
  {\doibase 10.1209/0295-5075/82/57003} {\bibfield  {journal} {\bibinfo
  {journal} {EPL (Europhysics Letters)}\ }\textbf {\bibinfo {volume} {82}},\
  \bibinfo {pages} {57003} (\bibinfo {year} {2008})}\BibitemShut {NoStop}%
\bibitem [{\citenamefont {Gull}, \citenamefont {Reichman},\ and\ \citenamefont
  {Millis}(2010)}]{Gull2010Boldlinediagrammatic}%
  \BibitemOpen
  \bibfield  {author} {\bibinfo {author} {\bibfnamefont {E.}~\bibnamefont
  {Gull}}, \bibinfo {author} {\bibfnamefont {D.~R.}\ \bibnamefont {Reichman}},
  \ and\ \bibinfo {author} {\bibfnamefont {A.~J.}\ \bibnamefont {Millis}},\
  }\href {\doibase 10.1103/PhysRevB.82.075109} {\bibfield  {journal} {\bibinfo
  {journal} {Phys. Rev. B}\ }\textbf {\bibinfo {volume} {82}},\ \bibinfo
  {pages} {075109} (\bibinfo {year} {2010})}\BibitemShut {NoStop}%
\bibitem [{\citenamefont {Gull}\ \emph {et~al.}(2011)\citenamefont {Gull},
  \citenamefont {Millis}, \citenamefont {Lichtenstein}, \citenamefont
  {Rubtsov}, \citenamefont {Troyer},\ and\ \citenamefont
  {Werner}}]{Gull2011ContinuoustimeMonte}%
  \BibitemOpen
  \bibfield  {author} {\bibinfo {author} {\bibfnamefont {E.}~\bibnamefont
  {Gull}}, \bibinfo {author} {\bibfnamefont {A.~J.}\ \bibnamefont {Millis}},
  \bibinfo {author} {\bibfnamefont {A.~I.}\ \bibnamefont {Lichtenstein}},
  \bibinfo {author} {\bibfnamefont {A.~N.}\ \bibnamefont {Rubtsov}}, \bibinfo
  {author} {\bibfnamefont {M.}~\bibnamefont {Troyer}}, \ and\ \bibinfo {author}
  {\bibfnamefont {P.}~\bibnamefont {Werner}},\ }\href@noop {} {\bibfield
  {journal} {\bibinfo  {journal} {Rev. Mod. Phys.}\ }\textbf {\bibinfo {volume}
  {83}},\ \bibinfo {pages} {349} (\bibinfo {year} {2011})},\ \Eprint
  {http://arxiv.org/abs/1012.4474} {1012.4474} \BibitemShut {NoStop}%
\bibitem [{\citenamefont {Werner}, \citenamefont {Oka},\ and\ \citenamefont
  {Millis}(2009)}]{Werner2009DiagrammaticMonteCarlo}%
  \BibitemOpen
  \bibfield  {author} {\bibinfo {author} {\bibfnamefont {P.}~\bibnamefont
  {Werner}}, \bibinfo {author} {\bibfnamefont {T.}~\bibnamefont {Oka}}, \ and\
  \bibinfo {author} {\bibfnamefont {A.~J.}\ \bibnamefont {Millis}},\ }\href
  {\doibase 10.1103/PhysRevB.79.035320} {\bibfield  {journal} {\bibinfo
  {journal} {Phys. Rev. B}\ }\textbf {\bibinfo {volume} {79}},\ \bibinfo
  {pages} {035320} (\bibinfo {year} {2009})}\BibitemShut {NoStop}%
\bibitem [{\citenamefont {Werner}\ and\ \citenamefont
  {Millis}(2010)}]{Werner2010Dynamicalscreeningin}%
  \BibitemOpen
  \bibfield  {author} {\bibinfo {author} {\bibfnamefont {P.}~\bibnamefont
  {Werner}}\ and\ \bibinfo {author} {\bibfnamefont {A.~J.}\ \bibnamefont
  {Millis}},\ }\href {\doibase 10.1103/PhysRevLett.104.146401} {\bibfield
  {journal} {\bibinfo  {journal} {Phys. Rev. Lett.}\ }\textbf {\bibinfo
  {volume} {104}},\ \bibinfo {pages} {146401} (\bibinfo {year}
  {2010})}\BibitemShut {NoStop}%
\bibitem [{\citenamefont {Cohen}\ and\ \citenamefont
  {Rabani}(2011)}]{Cohen2011Memoryeffectsin}%
  \BibitemOpen
  \bibfield  {author} {\bibinfo {author} {\bibfnamefont {G.}~\bibnamefont
  {Cohen}}\ and\ \bibinfo {author} {\bibfnamefont {E.}~\bibnamefont {Rabani}},\
  }\href@noop {} {\bibfield  {journal} {\bibinfo  {journal} {Phys. Rev. B}\
  }\textbf {\bibinfo {volume} {84}},\ \bibinfo {pages} {075150} (\bibinfo
  {year} {2011})}\BibitemShut {NoStop}%
\bibitem [{\citenamefont {Cohen}, \citenamefont {Wilner},\ and\ \citenamefont
  {Rabani}(2013)}]{Cohen2013Generalizedprojecteddynamics}%
  \BibitemOpen
  \bibfield  {author} {\bibinfo {author} {\bibfnamefont {G.}~\bibnamefont
  {Cohen}}, \bibinfo {author} {\bibfnamefont {E.~Y.}\ \bibnamefont {Wilner}}, \
  and\ \bibinfo {author} {\bibfnamefont {E.}~\bibnamefont {Rabani}},\
  }\href@noop {} {\bibfield  {journal} {\bibinfo  {journal} {New Journal of
  Physics}\ }\textbf {\bibinfo {volume} {15}},\ \bibinfo {pages} {073018}
  (\bibinfo {year} {2013})}\BibitemShut {NoStop}%
\bibitem [{\citenamefont {Cohen}\ \emph {et~al.}(2013)\citenamefont {Cohen},
  \citenamefont {Gull}, \citenamefont {Reichman}, \citenamefont {Millis},\ and\
  \citenamefont {Rabani}}]{Cohen2013Numericallyexactlong}%
  \BibitemOpen
  \bibfield  {author} {\bibinfo {author} {\bibfnamefont {G.}~\bibnamefont
  {Cohen}}, \bibinfo {author} {\bibfnamefont {E.}~\bibnamefont {Gull}},
  \bibinfo {author} {\bibfnamefont {D.~R.}\ \bibnamefont {Reichman}}, \bibinfo
  {author} {\bibfnamefont {A.~J.}\ \bibnamefont {Millis}}, \ and\ \bibinfo
  {author} {\bibfnamefont {E.}~\bibnamefont {Rabani}},\ }\href@noop {}
  {\bibfield  {journal} {\bibinfo  {journal} {Phys. Rev. B}\ }\textbf {\bibinfo
  {volume} {87}},\ \bibinfo {pages} {195108} (\bibinfo {year}
  {2013})}\BibitemShut {NoStop}%
\bibitem [{\citenamefont {Egger}\ and\ \citenamefont
  {Weiss}(1992)}]{Egger1992QuantumMonteCarlo}%
  \BibitemOpen
  \bibfield  {author} {\bibinfo {author} {\bibfnamefont {R.}~\bibnamefont
  {Egger}}\ and\ \bibinfo {author} {\bibfnamefont {U.}~\bibnamefont {Weiss}},\
  }\href {\doibase 10.1007/BF01320834} {\bibfield  {journal} {\bibinfo
  {journal} {Zeitschrift f{\"{u}}r Phys. B Condens. Matter}\ }\textbf {\bibinfo
  {volume} {89}},\ \bibinfo {pages} {97} (\bibinfo {year} {1992})}\BibitemShut
  {NoStop}%
\bibitem [{\citenamefont {Egger}\ and\ \citenamefont
  {Mak}(1994)}]{Egger1994Lowtemperaturedynamical}%
  \BibitemOpen
  \bibfield  {author} {\bibinfo {author} {\bibfnamefont {R.}~\bibnamefont
  {Egger}}\ and\ \bibinfo {author} {\bibfnamefont {C.~H.}\ \bibnamefont
  {Mak}},\ }\href {\doibase 10.1103/PhysRevB.50.15210} {\bibfield  {journal}
  {\bibinfo  {journal} {Phys. Rev. B}\ }\textbf {\bibinfo {volume} {50}},\
  \bibinfo {pages} {210} (\bibinfo {year} {1994})}\BibitemShut {NoStop}%
\bibitem [{\citenamefont {M{\"{u}}hlbacher}\ and\ \citenamefont
  {Egger}(2003)}]{Muehlbacher2003Crossoverfromnonadiabatic}%
  \BibitemOpen
  \bibfield  {author} {\bibinfo {author} {\bibfnamefont {L.}~\bibnamefont
  {M{\"{u}}hlbacher}}\ and\ \bibinfo {author} {\bibfnamefont {R.}~\bibnamefont
  {Egger}},\ }\href {\doibase 10.1063/1.1523014} {\bibfield  {journal}
  {\bibinfo  {journal} {J. Chem. Phys.}\ }\textbf {\bibinfo {volume} {118}},\
  \bibinfo {pages} {179} (\bibinfo {year} {2003})}\BibitemShut {NoStop}%
\bibitem [{\citenamefont {Cohen}\ \emph
  {et~al.}(2014{\natexlab{a}})\citenamefont {Cohen}, \citenamefont {Gull},
  \citenamefont {Reichman},\ and\ \citenamefont
  {Millis}}]{Cohen2014Greensfunctionsfrom}%
  \BibitemOpen
  \bibfield  {author} {\bibinfo {author} {\bibfnamefont {G.}~\bibnamefont
  {Cohen}}, \bibinfo {author} {\bibfnamefont {E.}~\bibnamefont {Gull}},
  \bibinfo {author} {\bibfnamefont {D.~R.}\ \bibnamefont {Reichman}}, \ and\
  \bibinfo {author} {\bibfnamefont {A.~J.}\ \bibnamefont {Millis}},\
  }\href@noop {} {\bibfield  {journal} {\bibinfo  {journal} {Phys. Rev. Lett.}\
  }\textbf {\bibinfo {volume} {112}},\ \bibinfo {pages} {146802} (\bibinfo
  {year} {2014}{\natexlab{a}})}\BibitemShut {NoStop}%
\bibitem [{\citenamefont {Cohen}\ \emph
  {et~al.}(2014{\natexlab{b}})\citenamefont {Cohen}, \citenamefont {Reichman},
  \citenamefont {Millis},\ and\ \citenamefont
  {Gull}}]{Cohen2014Greensfunctionsfroma}%
  \BibitemOpen
  \bibfield  {author} {\bibinfo {author} {\bibfnamefont {G.}~\bibnamefont
  {Cohen}}, \bibinfo {author} {\bibfnamefont {D.~R.}\ \bibnamefont {Reichman}},
  \bibinfo {author} {\bibfnamefont {A.~J.}\ \bibnamefont {Millis}}, \ and\
  \bibinfo {author} {\bibfnamefont {E.}~\bibnamefont {Gull}},\ }\href@noop {}
  {\bibfield  {journal} {\bibinfo  {journal} {Phys. Rev. B}\ }\textbf {\bibinfo
  {volume} {89}},\ \bibinfo {pages} {115139} (\bibinfo {year}
  {2014}{\natexlab{b}})}\BibitemShut {NoStop}%
\bibitem [{\citenamefont {Cohen}\ \emph {et~al.}(2015)\citenamefont {Cohen},
  \citenamefont {Gull}, \citenamefont {Reichman},\ and\ \citenamefont
  {Millis}}]{Cohen2015Tamingdynamicalsign}%
  \BibitemOpen
  \bibfield  {author} {\bibinfo {author} {\bibfnamefont {G.}~\bibnamefont
  {Cohen}}, \bibinfo {author} {\bibfnamefont {E.}~\bibnamefont {Gull}},
  \bibinfo {author} {\bibfnamefont {D.~R.}\ \bibnamefont {Reichman}}, \ and\
  \bibinfo {author} {\bibfnamefont {A.~J.}\ \bibnamefont {Millis}},\ }\href
  {\doibase 10.1103/PhysRevLett.115.266802} {\bibfield  {journal} {\bibinfo
  {journal} {Phys. Rev. Lett.}\ }\textbf {\bibinfo {volume} {115}},\ \bibinfo
  {pages} {266802} (\bibinfo {year} {2015})}\BibitemShut {NoStop}%
\bibitem [{\citenamefont {Leggett}\ \emph {et~al.}(1987)\citenamefont
  {Leggett}, \citenamefont {Chakravarty}, \citenamefont {Dorsey}, \citenamefont
  {Fisher}, \citenamefont {Garg},\ and\ \citenamefont
  {Zwerger}}]{Leggett1987Dynamicsdissipativetwo}%
  \BibitemOpen
  \bibfield  {author} {\bibinfo {author} {\bibfnamefont {A.~J.}\ \bibnamefont
  {Leggett}}, \bibinfo {author} {\bibfnamefont {S.}~\bibnamefont
  {Chakravarty}}, \bibinfo {author} {\bibfnamefont {A.~T.~A.}\ \bibnamefont
  {Dorsey}}, \bibinfo {author} {\bibfnamefont {M.~P.~A.}\ \bibnamefont
  {Fisher}}, \bibinfo {author} {\bibfnamefont {A.}~\bibnamefont {Garg}}, \ and\
  \bibinfo {author} {\bibfnamefont {W.}~\bibnamefont {Zwerger}},\ }\href
  {\doibase 10.1103/RevModPhys.59.1} {\bibfield  {journal} {\bibinfo  {journal}
  {Rev. Mod. Phys.}\ }\textbf {\bibinfo {volume} {59}},\ \bibinfo {pages} {1}
  (\bibinfo {year} {1987})}\BibitemShut {NoStop}%
\bibitem [{\citenamefont {Weiss}(1999)}]{Weiss1999Quantumdissipativesystems}%
  \BibitemOpen
  \bibfield  {author} {\bibinfo {author} {\bibfnamefont {U.}~\bibnamefont
  {Weiss}},\ }\href@noop {} {\emph {\bibinfo {title} {{Quantum dissipative
  systems}}}},\ Vol.~\bibinfo {volume} {10}\ (\bibinfo  {publisher} {World
  Scientific Publishing Company Incorporated},\ \bibinfo {year}
  {1999})\BibitemShut {NoStop}%
\bibitem [{\citenamefont {Nitzan}(2006)}]{Nitzan2006ChemicalDynamicsin}%
  \BibitemOpen
  \bibfield  {author} {\bibinfo {author} {\bibfnamefont {A.}~\bibnamefont
  {Nitzan}},\ }\href {\doibase 10.1002/cphc.200700074} {\emph {\bibinfo {title}
  {{Chemical Dynamics in Condensed Phases: Relaxation, Transfer, and Reactions
  in Condensed Molecular Systems}}}}\ (\bibinfo  {publisher} {Oxford University
  Press, New York},\ \bibinfo {year} {2006})\BibitemShut {NoStop}%
\bibitem [{\citenamefont {Wang}, \citenamefont {Thoss},\ and\ \citenamefont
  {Miller}(2001)}]{Wang2001Systematicconvergencein}%
  \BibitemOpen
  \bibfield  {author} {\bibinfo {author} {\bibfnamefont {H.}~\bibnamefont
  {Wang}}, \bibinfo {author} {\bibfnamefont {M.}~\bibnamefont {Thoss}}, \ and\
  \bibinfo {author} {\bibfnamefont {W.~H.}\ \bibnamefont {Miller}},\ }\href
  {\doibase 10.1063/1.1385561} {\bibfield  {journal} {\bibinfo  {journal} {J.
  Chem. Phys.}\ }\textbf {\bibinfo {volume} {115}},\ \bibinfo {pages} {2979}
  (\bibinfo {year} {2001})}\BibitemShut {NoStop}%
\bibitem [{\citenamefont {Thoss}, \citenamefont {Wang},\ and\ \citenamefont
  {Miller}(2001)}]{Thoss2001Selfconsistenthybrid}%
  \BibitemOpen
  \bibfield  {author} {\bibinfo {author} {\bibfnamefont {M.}~\bibnamefont
  {Thoss}}, \bibinfo {author} {\bibfnamefont {H.}~\bibnamefont {Wang}}, \ and\
  \bibinfo {author} {\bibfnamefont {W.~H.}\ \bibnamefont {Miller}},\ }\href
  {\doibase 10.1063/1.1385562} {\bibfield  {journal} {\bibinfo  {journal} {J.
  Chem. Phys.}\ }\textbf {\bibinfo {volume} {115}},\ \bibinfo {pages} {2991}
  (\bibinfo {year} {2001})}\BibitemShut {NoStop}%
\bibitem [{\citenamefont {Wang}\ and\ \citenamefont
  {Thoss}(2003)}]{Wang2003Multilayerformulationmulticonfiguration}%
  \BibitemOpen
  \bibfield  {author} {\bibinfo {author} {\bibfnamefont {H.}~\bibnamefont
  {Wang}}\ and\ \bibinfo {author} {\bibfnamefont {M.}~\bibnamefont {Thoss}},\
  }\href {\doibase 10.1063/1.1580111} {\bibfield  {journal} {\bibinfo
  {journal} {J. Chem. Phys.}\ }\textbf {\bibinfo {volume} {119}},\ \bibinfo
  {pages} {1289} (\bibinfo {year} {2003})}\BibitemShut {NoStop}%
\bibitem [{\citenamefont {Mak}\ and\ \citenamefont
  {Egger}(2007)}]{Mak2007MonteCarloMethods}%
  \BibitemOpen
  \bibfield  {author} {\bibinfo {author} {\bibfnamefont {C.~H.}\ \bibnamefont
  {Mak}}\ and\ \bibinfo {author} {\bibfnamefont {R.}~\bibnamefont {Egger}},\
  }\href {\doibase 10.1002/9780470141526.ch2} {\emph {\bibinfo {title} {Adv.
  Chem. Physics, New Methods Comput. Quantum Mech.}}},\ Vol.\ \bibinfo {volume}
  {XCIII}\ (\bibinfo  {publisher} {John Wiley {\&} Sons, Inc.},\ \bibinfo
  {year} {2007})\ pp.\ \bibinfo {pages} {39--76}\BibitemShut {NoStop}%
\bibitem [{\citenamefont {Makarov}\ and\ \citenamefont
  {Makri}(1994)}]{Makarov1994Pathintegralsdissipative}%
  \BibitemOpen
  \bibfield  {author} {\bibinfo {author} {\bibfnamefont {D.~E.}\ \bibnamefont
  {Makarov}}\ and\ \bibinfo {author} {\bibfnamefont {N.}~\bibnamefont
  {Makri}},\ }\href {\doibase 10.1016/0009-2614(94)00275-4} {\bibfield
  {journal} {\bibinfo  {journal} {Chem. Phys. Lett.}\ }\textbf {\bibinfo
  {volume} {221}},\ \bibinfo {pages} {482} (\bibinfo {year}
  {1994})}\BibitemShut {NoStop}%
\bibitem [{\citenamefont {Makri}(1995)}]{Makri1995Numericalpathintegral}%
  \BibitemOpen
  \bibfield  {author} {\bibinfo {author} {\bibfnamefont {N.}~\bibnamefont
  {Makri}},\ }\href {\doibase 10.1063/1.531046} {\bibfield  {journal} {\bibinfo
   {journal} {J. Math. Phys.}\ }\textbf {\bibinfo {volume} {36}},\ \bibinfo
  {pages} {2430} (\bibinfo {year} {1995})}\BibitemShut {NoStop}%
\bibitem [{\citenamefont {Tanimura}\ and\ \citenamefont
  {Kubo}(1989)}]{Tanimura1989TimeEvolutionQuantum}%
  \BibitemOpen
  \bibfield  {author} {\bibinfo {author} {\bibfnamefont {Y.}~\bibnamefont
  {Tanimura}}\ and\ \bibinfo {author} {\bibfnamefont {R.}~\bibnamefont
  {Kubo}},\ }\href {\doibase 10.1143/JPSJ.58.101} {\bibfield  {journal}
  {\bibinfo  {journal} {Journal of the Physical Society of Japan}\ }\textbf
  {\bibinfo {volume} {58}},\ \bibinfo {pages} {101} (\bibinfo {year}
  {1989})}\BibitemShut {NoStop}%
\bibitem [{\citenamefont {Str{\"{u}}mpfer}\ and\ \citenamefont
  {Schulten}(2012)}]{Struempfer2012OpenQuantumDynamics}%
  \BibitemOpen
  \bibfield  {author} {\bibinfo {author} {\bibfnamefont {J.}~\bibnamefont
  {Str{\"{u}}mpfer}}\ and\ \bibinfo {author} {\bibfnamefont {K.}~\bibnamefont
  {Schulten}},\ }\href {\doibase 10.1021/ct3003833} {\bibfield  {journal}
  {\bibinfo  {journal} {J. Chem. Theory Comput.}\ }\textbf {\bibinfo {volume}
  {8}},\ \bibinfo {pages} {2808} (\bibinfo {year} {2012})}\BibitemShut
  {NoStop}%
\bibitem [{\citenamefont
  {Van~Kampen}(1974)}]{VanKampen1974cumulantexpansionstochastic}%
  \BibitemOpen
  \bibfield  {author} {\bibinfo {author} {\bibfnamefont {N.}~\bibnamefont
  {Van~Kampen}},\ }\href@noop {} {\bibfield  {journal} {\bibinfo  {journal}
  {Physica}\ }\textbf {\bibinfo {volume} {74}},\ \bibinfo {pages} {215}
  (\bibinfo {year} {1974})}\BibitemShut {NoStop}%
\bibitem [{\citenamefont {Yoon}, \citenamefont {Deutch},\ and\ \citenamefont
  {Freed}(1975)}]{Yoon1975comparisongeneralizedcumulant}%
  \BibitemOpen
  \bibfield  {author} {\bibinfo {author} {\bibfnamefont {B.}~\bibnamefont
  {Yoon}}, \bibinfo {author} {\bibfnamefont {J.}~\bibnamefont {Deutch}}, \ and\
  \bibinfo {author} {\bibfnamefont {J.~H.}\ \bibnamefont {Freed}},\ }\href@noop
  {} {\bibfield  {journal} {\bibinfo  {journal} {J. Chem. Phys.}\ }\textbf
  {\bibinfo {volume} {62}},\ \bibinfo {pages} {4687} (\bibinfo {year}
  {1975})}\BibitemShut {NoStop}%
\bibitem [{\citenamefont {Mukamel}(1979)}]{Mukamel1979NonMarkoviantheory}%
  \BibitemOpen
  \bibfield  {author} {\bibinfo {author} {\bibfnamefont {S.}~\bibnamefont
  {Mukamel}},\ }\href@noop {} {\bibfield  {journal} {\bibinfo  {journal} {Chem.
  Phys.}\ }\textbf {\bibinfo {volume} {37}},\ \bibinfo {pages} {33} (\bibinfo
  {year} {1979})}\BibitemShut {NoStop}%
\bibitem [{\citenamefont {Reichman}, \citenamefont {Brown},\ and\ \citenamefont
  {Neu}(1997)}]{Reichman1997Cumulantexpansionsand}%
  \BibitemOpen
  \bibfield  {author} {\bibinfo {author} {\bibfnamefont {D.~R.}\ \bibnamefont
  {Reichman}}, \bibinfo {author} {\bibfnamefont {F.~L.~H.}\ \bibnamefont
  {Brown}}, \ and\ \bibinfo {author} {\bibfnamefont {P.}~\bibnamefont {Neu}},\
  }\href {\doibase 10.1103/PhysRevE.55.2328} {\bibfield  {journal} {\bibinfo
  {journal} {Phys. Rev. E}\ }\textbf {\bibinfo {volume} {55}},\ \bibinfo
  {pages} {2328} (\bibinfo {year} {1997})}\BibitemShut {NoStop}%
\bibitem [{\citenamefont {Metropolis}\ \emph {et~al.}(1953)\citenamefont
  {Metropolis}, \citenamefont {Rosenbluth}, \citenamefont {Rosenbluth},
  \citenamefont {Teller},\ and\ \citenamefont
  {Teller}}]{Metropolis1953EquationStateCalculations}%
  \BibitemOpen
  \bibfield  {author} {\bibinfo {author} {\bibfnamefont {N.}~\bibnamefont
  {Metropolis}}, \bibinfo {author} {\bibfnamefont {A.~W.}\ \bibnamefont
  {Rosenbluth}}, \bibinfo {author} {\bibfnamefont {M.~N.}\ \bibnamefont
  {Rosenbluth}}, \bibinfo {author} {\bibfnamefont {A.~H.}\ \bibnamefont
  {Teller}}, \ and\ \bibinfo {author} {\bibfnamefont {E.}~\bibnamefont
  {Teller}},\ }\href {\doibase 10.1063/1.1699114} {\bibfield  {journal}
  {\bibinfo  {journal} {J. Chem. Phys.}\ }\textbf {\bibinfo {volume} {21}},\
  \bibinfo {pages} {1087} (\bibinfo {year} {1953})}\BibitemShut {NoStop}%
\bibitem [{\citenamefont {Hastings}(1970)}]{Hastings1970MonteCarlosampling}%
  \BibitemOpen
  \bibfield  {author} {\bibinfo {author} {\bibfnamefont {W.~K.}\ \bibnamefont
  {Hastings}},\ }\href {\doibase 10.1093/biomet/57.1.97} {\bibfield  {journal}
  {\bibinfo  {journal} {Biometrika}\ }\textbf {\bibinfo {volume} {57}},\
  \bibinfo {pages} {97} (\bibinfo {year} {1970})}\BibitemShut {NoStop}%
\bibitem [{\citenamefont {Antipov}, \citenamefont {Dong},\ and\ \citenamefont
  {Gull}(2016)}]{Antipov2016VoltageQuenchDynamics}%
  \BibitemOpen
  \bibfield  {author} {\bibinfo {author} {\bibfnamefont {A.~E.}\ \bibnamefont
  {Antipov}}, \bibinfo {author} {\bibfnamefont {Q.}~\bibnamefont {Dong}}, \
  and\ \bibinfo {author} {\bibfnamefont {E.}~\bibnamefont {Gull}},\ }\href
  {\doibase 10.1103/PhysRevLett.116.036801} {\bibfield  {journal} {\bibinfo
  {journal} {Phys. Rev. Lett.}\ }\textbf {\bibinfo {volume} {116}},\ \bibinfo
  {pages} {036801} (\bibinfo {year} {2016})}\BibitemShut {NoStop}%
\bibitem [{Note1()}]{Note1}%
  \BibitemOpen
  \bibinfo {note} {In our calculations in the companion paper, we compare the
  full error estimation and the statistical errors from individual runs,
  showing that they follow a nontrivial relationship. It is possible to perform
  a full error analysis within a single Monte Carlo run by way of a recursive
  bootstrapping procedure; however, this has not been implemented so far, and
  it is not clear that it offers an advantage.}\BibitemShut {Stop}%
\bibitem [{\citenamefont {Fetter}\ and\ \citenamefont
  {Walecka}(2003)}]{Fetter2003QuantumTheoryMany}%
  \BibitemOpen
  \bibfield  {author} {\bibinfo {author} {\bibfnamefont {A.~L.}\ \bibnamefont
  {Fetter}}\ and\ \bibinfo {author} {\bibfnamefont {J.~D.}\ \bibnamefont
  {Walecka}},\ }\href@noop {} {\emph {\bibinfo {title} {Quantum Theory of
  Many-Particle Systems}}}\ (\bibinfo  {publisher} {Courier Corporation},\
  \bibinfo {year} {2003})\ \bibinfo {note} {google-Books-ID:
  0wekf1s83b0C}\BibitemShut {NoStop}%
\bibitem [{\citenamefont {Werner}\ and\ \citenamefont
  {Millis}(2007)}]{Werner2007Efficientdynamicalmean}%
  \BibitemOpen
  \bibfield  {author} {\bibinfo {author} {\bibfnamefont {P.}~\bibnamefont
  {Werner}}\ and\ \bibinfo {author} {\bibfnamefont {A.~J.}\ \bibnamefont
  {Millis}},\ }\href {\doibase 10.1103/PhysRevLett.99.146404} {\bibfield
  {journal} {\bibinfo  {journal} {Phys. Rev. Lett.}\ }\textbf {\bibinfo
  {volume} {99}},\ \bibinfo {pages} {146404} (\bibinfo {year}
  {2007})}\BibitemShut {NoStop}%
\bibitem [{\citenamefont {Werner}\ and\ \citenamefont
  {Eckstein}(2013)}]{Werner2013Phononenhancedrelaxation}%
  \BibitemOpen
  \bibfield  {author} {\bibinfo {author} {\bibfnamefont {P.}~\bibnamefont
  {Werner}}\ and\ \bibinfo {author} {\bibfnamefont {M.}~\bibnamefont
  {Eckstein}},\ }\href {\doibase 10.1103/PhysRevB.88.165108} {\bibfield
  {journal} {\bibinfo  {journal} {Phys. Rev. B}\ }\textbf {\bibinfo {volume}
  {88}},\ \bibinfo {pages} {165108} (\bibinfo {year} {2013})}\BibitemShut
  {NoStop}%
\bibitem [{\citenamefont {Chen}\ \emph {et~al.}(2016)\citenamefont {Chen},
  \citenamefont {Cohen}, \citenamefont {Millis},\ and\ \citenamefont
  {Reichman}}]{chen_anderson-holstein_2016}%
  \BibitemOpen
  \bibfield  {author} {\bibinfo {author} {\bibfnamefont {H.-T.}\ \bibnamefont
  {Chen}}, \bibinfo {author} {\bibfnamefont {G.}~\bibnamefont {Cohen}},
  \bibinfo {author} {\bibfnamefont {A.~J.}\ \bibnamefont {Millis}}, \ and\
  \bibinfo {author} {\bibfnamefont {D.~R.}\ \bibnamefont {Reichman}},\ }\href
  {\doibase 10.1103/PhysRevB.93.174309} {\bibfield  {journal} {\bibinfo
  {journal} {Physical Review B}\ }\textbf {\bibinfo {volume} {93}},\ \bibinfo
  {pages} {174309} (\bibinfo {year} {2016})}\BibitemShut {NoStop}%
\bibitem [{\citenamefont {Profumo}\ \emph {et~al.}(2015)\citenamefont
  {Profumo}, \citenamefont {Groth}, \citenamefont {Messio}, \citenamefont
  {Parcollet},\ and\ \citenamefont {Waintal}}]{Profumo2015QuantumMonteCarlo}%
  \BibitemOpen
  \bibfield  {author} {\bibinfo {author} {\bibfnamefont {R.~E.}\ \bibnamefont
  {Profumo}}, \bibinfo {author} {\bibfnamefont {C.}~\bibnamefont {Groth}},
  \bibinfo {author} {\bibfnamefont {L.}~\bibnamefont {Messio}}, \bibinfo
  {author} {\bibfnamefont {O.}~\bibnamefont {Parcollet}}, \ and\ \bibinfo
  {author} {\bibfnamefont {X.}~\bibnamefont {Waintal}},\ }\href@noop {}
  {\bibfield  {journal} {\bibinfo  {journal} {Phys. Rev. B}\ }\textbf {\bibinfo
  {volume} {91}},\ \bibinfo {pages} {245154} (\bibinfo {year}
  {2015})}\BibitemShut {NoStop}%
\bibitem [{\citenamefont {Messiah}(1961)}]{Messiah1961QuantumMechanics}%
  \BibitemOpen
  \bibfield  {author} {\bibinfo {author} {\bibfnamefont {A.}~\bibnamefont
  {Messiah}},\ }\href@noop {} {\emph {\bibinfo {title} {Quantum {Mechanics}}}}\
  (\bibinfo  {publisher} {Dover Publications},\ \bibinfo {year}
  {1961})\BibitemShut {NoStop}%
\end{thebibliography}%

\end{document}